\documentclass[12pt,thmsa]{article}
\usepackage{amssymb}

\usepackage{sw20lart}

\input{tcilatex}
\begin{document}

\title{Stochastic interpretation of quantum mechanics assuming that vacuum fields
are real}
\author{Emilio Santos \\
Departamento de F\'{i}sica. Universidad de Cantabria. Santander. Spain}
\maketitle

\begin{abstract}
I review the realistic interpretation of several typically quantum phenomena
using a heuristic approach that rests on the assumption that the
electromagnetic quantum vacuum is a stochastic field. I include the particle
behaviour of light, the photoelectric effect, the hdrogen atom, the Casimir
effect, and entanglement in the optical tests of Bell inequalities. The
stochastic approach might be formally connected with the standard Hilbert
space formalism via the Wigner representation of the field.
\end{abstract}

\section{ Introduction}

One century after the discovery of quantum mechanics we still lack any
consensus about what one is actually talking about as one uses it. ``There
is a gap between the abstract terms in which the theory is couched and the
phenomena the theory enables each of us to account for so well. Because it
has no practical consequences for how we each use quantum mechanics to deal
with physical problems, this cognitive dissonance has managed to coexist
with the quantum theory from the very beginning''\cite{Mermin}.

The discrepancy about the correct approach to the theory appeared very
early, two extremes corresponding to the creators of `wave mechanics' (de
Broglie, Schr\"{o}dinger) and those of `quantum mechanics' (Heisenberg,
Bohr, Pauli). People in the former group attempted to get a picture of the
microworld, without real success. Those in the latter supported the view
that a picture of reality is not needed. The absence of a satisfactory
picture, in spite of a big effort by some people, combined with the
mathematical elegance of the (Hilbert space) formalism of quantum mechanics
plus its spectacular success in the quantitative predictions of empirical
evidence, led the mainstream of the community to support the Heisenberg-Bohr
view (the Copenhagen interpretation). In recent times alternative
interpretations have been proposed like `many worlds', bizarre in my
opinion, or explanations for the lack of consensus like QBism. The latter
rests on the belief that ``the absence of conceptual clarity for almost a
century suggests that the problem might lie in some implicit misconceptions
about the nature of scientific explanation'' \cite{Mermin}. In section 2 of
this article I will provide arguments for both the possibility and the
usefulness of a realistic interpretation of quantum mechanics. See also \cite
{book}.

I believe that in order to achieve a realistic interpretation we must assume
that quantum vacuum fields are real. On the other hand a plausible
explanation for the stability of atoms, without departing from Maxwell
theory, is the existence of a background radiation filling space. The
conjunction of the two facts suggests to identify the background radiation
with the quantum vacuum electromagnetic field, taken as a stochastic field.
Indeed relativistic invariance leads to the spectrum of the possible
radiation, modulo a unique parameter fixing the scale. If we identify that
parameter with Planck constant there is agreement between the properties of
the assumed background radiation and the quantum vacuum field. Section 3 of
the article provides a more detailed exposition of this idea.

In section 4 we study the characterization of the vacuum electromagnetic
radiation as a stochastic field. Section 5 deals with particle properties of
light explained as due to the action of the stochastic vacuum fields. The
quantitative conection of the stochastic approach with the standard quantum
Hilbert space formalism is made via de Weyl transform leading to the Winger
representation, which is studied in section 6. Section 7 shows that the
vacuum fields allows understanting entanglement in experiments of
correlation between ``photon pairs'', e.g. in optical tests of Bell
inequalitis. Finally in section 8 I offer several ideas for the search of a
more complete realistic interpretation of quantum theory.

\section{Understanding vs. using quantum mechanics}

\subsection{Pragmatic approach}

None of the interpretations of quantum mechanics proposed till now\cite
{Drummond} offer a clear intuitive picture of the quantum world.
Nevertheless most physicists do not worry for the lack of a picture and
embrace a \textit{pragmatic approach} close to the early proposal of Bohr
and Heisenberg, usually known as the Copenhagen intepretation\cite{Isham}.

Behind the pragmatic approach there is usually a philosophical position
about physics (or science in general) that may be summarized as follows. It
is taken for granted that a physical theory has at least two components\cite
{Suppe}: (1) the formalism, or mathematical apparatus, of the theory, and
(2) the correspondence rules that establish a link between the formalism and
the results of observations or measurements. As an example let us consider
the formalism of quantum mechanics based on the mathematical theory of
Hilbert spaces. The formalism involves two kinds of operators, density
operators, $\hat{\rho},$ that represent states, and self-adjoint operators, $%
\hat{A}$, that represent observables. The link with the measurement results
is given by the postulate that the expectation value, $Tr\left( \hat{\rho}%
\hat{A}\right) ,$ corresponds to the statistical mean of the values obtained
when one realizes several measurements on identically prepared systems
(which determines $\hat{\rho})$ by means of an appropriate apparatus (that
corresponds to $\hat{A}).$

If we assume that the formalism and the correspondence rules are the \textit{%
only objects required to define a physical theory}, in the sense that the
statistical regularities need not be further explained, then we get what has
been called a \textit{minimal instrumentalistic interpretation} of the theory%
\cite{Redhead},\cite{Isham}. It may be identified with the purely pragmatic
approach mentioned above.

Most people claiming to support that approach accept the following positions:

1. The notion of an individual physical system `having' or `possessing'
values for all its physical quantities is \textit{inappropriate} in the
context of quantum theory.

2. The concept of `measurement' is fundamental in the sense that the scope
of quantum theory is \textit{intrinsecally} restricted to predicting the
results of measurements.

3. The spread in the results of measurements on identically prepared systems
must not be interpreted as reflecting a `lack of knowledge' of some
objectively existing state of affears.

The \textit{instrumentalistic} approach is quite different from, even
opposite to, the \emph{realistic} view traditional in classical physics.
Between these two extremes there are a variety of approaches.

\subsection{Realistic interpretation}

The main oponent to the purely pragmatic approach to quantum mechanics was
Albert Einstein. Indeed his discussions with Niels Bohr are the paradigm of
a scientific debate, hard in the scientific arguments but hearty from the
personal point of view. One of the most celebrated moments of the debate was
a 1935 article by Einstein, Podolsky and Rosen\cite{EPR} (EPR). It begins as
follows: ``Any serious consideration of a physical theory must take into
account the distinction between the objective reality, which is independent
of any theory, and the physical concepts with which the theory operates.
These concepts are intended to correspond with \textit{the objective reality}%
, and by means of these concepts \textit{we picture this reality to ourselves%
}'' (my emphasis).

I strongly support Einstein's view, that is I believe that a realistic
interpretation is possible. The main point is the claim that any physical
theory should offer a \emph{physical model} in addition to \emph{the
formalism and rules for the connection with the experiments. }The latter are
obviously essential because they are required for the comparison of the
theory with empirical evidence, which is \emph{the test} for the validity of
the theory. In my opinion physical models are also necessary in order to 
\emph{reach a coherent picture }of\emph{\ }the world. Many quantum
physicists apparently support the useless of pictures, but it is the case
that when they attempt popular explanations of quantum phenomena they
frequently propose actual pictures, many of them rather bizarre. For
instance it has been claimed that quantum mechanics \emph{compel us } to
believe that there are a multiplicity of `me' in parallel universes (the
many worlds interpretation) or that an atom may be present in two distant
places at the same time. This is an indication that the need of ``picture
the reality to ourselves''\cite{EPR} cannot be easily dismissed. Furthermore
the existence of physical models might open the possibility for new
developments and applications of quantum theory and therefore it is not a
purely academic question.

An illuminating confrontation between pragmatic and realistic epistemologies
is the conversation of Heisenberg with Einstein that took place in Berlin
1926, as remembered by Heisenberg himself\cite{Heisenberg26}. The most
relevant part is reproduced in the following:

``Einstein opened the conversation with a question that bore on the
philosophical background of my recent work.`What you have told us sounds
extremely strange. You assume the existence of electrons inside the atom,
and you are probably quite right to do so. But you refuse to consider their
orbits, even though we can observe electron tracks in a cloudchamber. I
should very much like to hear more about your reasons for making such
strange assumptions'. `We cannot observe electron orbits inside the atom', I
must have repIied, `but the radiation which an atom emits during discharges
enables us to deduce the frequencies and corresponding amplitudes of its
electrons. After all, even in the older physics wave numbers and amplitudes
could be considered substitutes for electron orbits. Now, since \textit{a
good theory must be based on directly observable magnitudes}, I thought it
more fitting to restrict myself to these, treating them, as it were, as
representatives of the electron orbits.' `But you don't seriously believe',
Einstein protested, `that none but observable magnitudes must go into a
physical theory?'.

The conversation continued for a while and at the end Einstein warned: ``%
\textit{You are moving on very thin ice. For you are suddenly speaking of
what we know about nature and no longer about what nature really does. In
science we ought to be concerned solely with what nature does}'' (my
emphasis). Einstein arguments are a clear support to a realistic
epistomology, and I fully agree with his views about the foundations of
quantum physics.

I propose that the difficulties for a realistic interpretation of quantum
phenomena do not derive from the empirical facts, or not only. Nevertheless
most textbooks of quantum mechanics emphasize the difficulty, or
impossibility, to interpret typical quantum phenomena with a realistic view.
The purpose of this article is to show that in fact those phenomena are
compatible with a picture of the microworld. Of course the picture is
somewhat different from the one offered by classical physics but not
dramatically different.

\section{Vacuum fields, the clue for a realistic interpretation}

The belief that the vacuum is not empty has been supported by many people
from long ago. It goes back, at least, to the idea of the \textit{ether }in
19th Century, that apparently was excluded by relativity theory. However it
reappeared with the development of quantum theory. Thus for instance de
Broglie theory or the hydrodynamical interpretation of Schr\"{o}dinger
equation by Madelung suggest a vacuum not empty. This has led to many
attempts to derive quantum theory, or at least Schr\"{o}dinger equation,
from the existence of a subquantum fluid \cite{book}.

The main assumption in this article is that the vacuum consists of several
stochastic fields, which precisely correspond to the quantum vacuum fields.
Here I will study only the implications of the vacuum electromagnetic
radiation. In the following I show that the stability of matter compel us,
or strongly suggests, the existence of stochastic fields which may be
identified with the quantum vacuum. Then I will discuss two relevant
predictions of the vacuum electromagnetic field: the energy and size of
atoms and the Casimir effect.

\subsection{The stability of atoms rests on vacuum radiation}

Soon after Rutherford experiment of 1911 that lead to the nuclear atom, Bohr
proposed in 1913 a model that involved postulates contradicting classical
electrodynamics. The common wisdom was, and it is still, that the
contradiction cannot be avoided. That it appears even for the most basic
empirical fact, the stability of matter, in particular atoms. But this claim
is flawed\cite{Santos68}.

Indeed a hydrogen atom, consisting of one proton and one electron, cannot be
stable if studied within classical electrodynamics. However the conclusion
is correct only \textit{if the atom is isolated}. The reason is that an
electron moving around the proton would radiate, and therefore the atom will
loss energy until it collapses. But the argument is not valid if there are
many atoms in the universe, as is the case, because if all atoms radiate the
hypothesis of isolation is not appropriate. It is more plausible to assume
that there is some amount of radiation filling space. If this is the case
then some radiation will fill space so that every atom would sometimes
radiate but other times it would absorb energy from the radiation,
eventually arriving at a dynamical equilibrium. This may explain, at least
qualitatively, the stability of the atom. The moral is that \textit{matter
and radiation of the universe cannot be studied independently within
classical electrodynamics, and the complexity of the universe compel us to
treat the assumed radiation as a stochastic field}. Then the electron of a
hydrogen atom would move in a random way around the nucleus. I propose that
the probability distribution of electron positions is what the
Schr\"{o}dinger wavefunction provides via Born\'{}s rule.

\subsection{Spectrum of the vacuum radiation}

It is plausible that the statistical properties of the assumed background
radiation are homogeneous, isotropic and Lorentz invariant. The most
relevant statistical property of a noise is the spectrum, $S(\omega ),$
defined as the radiation energy per unit volume and unit frequency interval.
It is the case that the unique spectrum compatible with the said constraints
is a spectrum proportional to the cube of the frequency \cite{Dice}, that is

\begin{equation}
S(\omega )=const.\times \omega ^{3}.  \label{s}
\end{equation}
We shall fix the constant in order to fit empirical results, whence we will
assume 
\begin{equation}
S(\omega )=\frac{
\rlap{\protect\rule[1.1ex]{.325em}{.1ex}}h%
}{2\pi ^{2}c^{3}}\omega ^{3},  \label{01}
\end{equation}
where $
\rlap{\protect\rule[1.1ex]{.325em}{.1ex}}h%
$ is Planck constant.

A standard method to study the radiation field in free space is to expand it
in plane waves (or in normal modes if it is enclosed in a cavity). In free
space the number of modes, $N$, per unit volume and unit frequency interval
is 
\begin{equation}
N=\frac{\omega ^{2}}{\pi ^{2}c^{3}}.  \label{00}
\end{equation}
Taking eq.$\left( \ref{01}\right) $ into account the vacuum radiation field
is equivalent to an energy 
\begin{equation}
E=\frac{S}{N}=\frac{1}{2}
\rlap{\protect\rule[1.1ex]{.325em}{.1ex}}h%
\omega  \label{02}
\end{equation}
per normal mode of the radiation. Eq.$\left( \ref{02}\right) $ is just $1/2$
the ``quantum'' of energy introduced by Planck in his pioneer derivation of
the radiation law that gave birth to quantum theory. In the following I will
derive some consequences of the existence of vacuum radiation with spectrum
eq$.\left( \ref{01}\right) .$

\subsection{Motion under the action of the vacuum radiation}

A frequenlty quoted difference between classical and quantum physics is the
impossibility of measuring simultaneously the position and the velocity (or
momentum) of a particle in the quantum domain. But it is the case that the
impossibility is a straightforward consequence of stochasticity. In fact the
motion of a charged particle under the action of the vacuum radiation would
be irregular due to the action of the field. That is, the position may
change during short time intervals, something similar to what happens in
Brownian motion. This is general for measurements en the presence of noise,
but there are two peculiar properties due to the specific spectrum eq.$%
\left( \ref{01}\right) .$ Firstly the conservation of some memory of the
velocity during long times in the motion under not too strong external
forces, and second the Heisenberg uncertainty relations. The former property
is derived in the following, the latter in the next subsection.

The spectrum of the vacuum radiation, eq.$\left( \ref{s}\right) ,$ is small
for low frequencies but strong for high ones. This is quite different from
what happens in Brownian motion whose noise is white, that is the spectrum
fulfils $S(\omega )=$ constant$,$ independent of the frequency. With the
spectrum eq.$\left( \ref{s}\right) $ of the vacuum radiation, if the
velocity of a charged particle at the initial time $t=0$ is \textbf{v}$,$
then at time $T$ there will be a probability distribution of velocities with
an average 
\begin{equation}
\left\langle \mathbf{v}(T)\right\rangle =\left\langle \mathbf{v+}\frac{1}{m}%
\int_{0}^{T}\mathbf{F}(t)dt\right\rangle =\mathbf{v+}\frac{1}{m}%
\int_{0}^{T}\left\langle \mathbf{F}(t)\right\rangle dt\simeq \mathbf{v},
\label{Ft}
\end{equation}
where $\left\langle {}\right\rangle $ means average. The mean force $%
\left\langle \mathbf{F}(t)\right\rangle $ due to the vacuum field is nil
because that radiation is assumed isotropic, whence all directions of the
force would be equaly probable at any time. But the relevant point is that
the short time behaviour depends on the hight frequencies of the spectrum
while the long time depends on the low frequencies. Thus the spectrum eq.$%
\left( \ref{01}\right) $ gives rise to a short time shake because $S(\omega
) $ eq.$\left( \ref{s}\right) $ is large at high frequencies. However
preserves some memory during long times because $S(\omega )$ is small at low
frequencies. That is the latter equality in eq.$\left( \ref{Ft}\right) $
holds for relatively long times, in sharp contrast with Brownian motion
where the memory is completely lost. These facts fit in the quantum
mechanical prediction that the momentum is preserved in free particle
motion. In contrast to eq.$\left( \ref{Ft}\right) ,$ when the particle is
not free the evolution of the mean velocity is involved because the actions
of the sure force and the random one $\mathbf{F}(t)$ become entangled.

Eq.$\left( \ref{Ft}\right) $ requires that the particle is charged, but we
may assume that something similar happens for neutral particles with charged
parts.

\subsection{Heisenberg uncertainty relations}

Uncertainty relations are frequently quoted a one of the characteristic
traits of quantum mechanics, in contrast with classical physics. Here I
shall show that they are a consequence of the spectrum eq.$\left( \ref{01}%
\right) $ of the vacuum radiation. In fact let us consider a charged
particle moving in a deep enough potential well (in one dimension for
simplicity). The motion will be roughly periodic with some frequency $\omega
.$ It is plausible that the main interaction with the vacuum radiation takes
place via those normal modes of the field that have frequencies close to
those of the particle motion. Also the mean kinetic energy of the particle
should be close to one half the average energy of those normal modes which
have the greatest interaction with the atom. As the potential energy is of
order the kinetic energy with the sign changed. (The relation will be exact,
in view of the virial theorem, for a harmonic potential well. Then the total
energy should be the negative of the kinetic energy. That is we may write 
\begin{equation}
\left| E\right| =\frac{1}{2}mv^{2}\sim \frac{1}{2} 
\rlap{\protect\rule[1.1ex]{.325em}{.1ex}}h%
\omega  \label{meanv}
\end{equation}
Also the mean square velocity and the mean square position coordinate may be
related to the frequency as follows 
\begin{equation}
\left\langle v^{2}\right\rangle \sim \omega ^{2}\left\langle
x^{2}\right\rangle .  \label{meanx}
\end{equation}
Removing $\omega $ amongst these two equations we get

\[
m^{2}\left\langle v^{2}\right\rangle \left\langle x^{2}\right\rangle
=\left\langle x^{2}\right\rangle \left\langle p^{2}\right\rangle \sim 
\rlap{\protect\rule[1.1ex]{.325em}{.1ex}}h%
^{2}, 
\]
which roughly fits in Heisenberg uncertainty relation.

\subsection{The energy and size of the hydrogen atom}

Via an heuristic approach it is possible to derive the typical sizes and
energies of quantum systems governed by electromagnetic interactions. Let us
consider the hydrogen atom consisting of two particles, proton and electron,
characterized each by the mass and the electric charge. The proton mass
being much larger that the electron mass we may study the atom assuming that
the proton is at rest and the motion of the electron is such that the atom
is in a dynamical equilibrium with radiation. In our study of the electron
motion it is plausible that the main interaction with the vacuum radiation
takes place via those normal modes of the field that have frequencies close
to those of the electron motion. Also the mean kinetic energy of the
electron should be close to one half the average energy of those normal
modes which have the greatest interaction with the atom. As the potential
energy is twice the kinetic energy with the sign changed, in view of the
virial theorem, the total energy should be the negative of the kinetic
energy. Then if the electron moved around the nucleus in a circle having
energy $E$ (i. e. with balanced mean emission and absorption of radiation),
we might write the following equalities 
\begin{equation}
\left| E\right| =\frac{1}{2}mv^{2}=\frac{1}{2}\frac{e^{2}}{r},v=r\omega
,\left| E\right| \sim \frac{1}{2}
\rlap{\protect\rule[1.1ex]{.325em}{.1ex}}h%
\omega ,  \label{03}
\end{equation}
the latter corresponding to the condition of dynamical equilibrium with
radiation. Of course the motion is perturbed by the action of the vacuum
fields, whence the electron motion would be very irregular, not circular,
but it is plausible that eqs.$\left( \ref{03}\right) $ might be roughly
fulfilled on the average. Hence the energy and the size of the atom may be
got removing the quantities $v$ and $\omega $ from eqs.$\left( \ref{03}%
\right) $, which leads to 
\begin{equation}
E\sim -\frac{me^{4}}{2
\rlap{\protect\rule[1.1ex]{.325em}{.1ex}}h%
^{2}},r\sim \frac{
\rlap{\protect\rule[1.1ex]{.325em}{.1ex}}h%
^{2}}{me^{2}},  \label{04}
\end{equation}
in rough agreement with the quantum predictions and with experiments.

In this example we have used a heuristic approach, a rigorous stochastic
treatment would be more lengthy because it should involve also the vacuum
electron-positron field and possibly other fields, something that is not yet
available. It is remarkable that the standard quantum formalism allows a
relatively simple treatment, via Schr\"{o}dinger equation. The combination
of relative simplicity with a good agreement with empirical data is is the
great virtue of quantum mechanics. However it provides no picture of the
microscopic phenomena.

It is the case that the study of the vacuum electromagnetic radiation field
interacting, via Maxwell-Lorentz laws, with electric charges or macroscopic
bodies reproduces correctly several quantum predictions. Indeed extensive
research on this line has been made, which is known as stochastic, or
random, electrodynamics (SED). For reviews see \cite{Dice},\cite{Santos20}, 
\cite{book}. Actually there are also cases where the SED predictions
disagree with quantum mechanics (and experiments), a fact that we may
attribute to the neglect of: 1) other vacuum fields, like electron-positron,
and 2) the back action of the charges that would modify the vacuum
radiation. A case where the SED treatment fully agrees with the quantum one
is the Casimir effect that we briefly revisit below.

\subsection{Stationary states of charged particles. Bohr atomic model}

The existence of vacuum radiation suggests a physical picture for the energy
spectra of quantum systems, not only for their ground states. In particular
for the hydrogen atom as is shown in the following.

In spite of this we may assume that when a particle follows a path close to
classical, although suffering strong shaking, large deviations from the
classical orbit might be scarce. Thus in the hydrogen atom some orbits of
the electron that would be periodic according to classical mechanics would
be relatively stable. These orbits have constant angular momentum (they are
ellipses or in particular circles). It is not strange that these were the
orbits quantized in the Bohr-Sommerfeld model of the atom. The idea was the
basis of the ``old quantum theory'', where quantization was made in terms of
action-angle variables. Of course the old quantum theory is known to be a
semiclassical approximation to modern quantum mechanics.

In the following I will give arguments that may provide a physical picture
for the atimic Bohr model, which rests on two celebrated postulates. The
second one is just the assumption, proposed earlier by Planck, that
absorption of radiation takes place in the form of ``quanta'' with energy 
\begin{equation}
E=
\rlap{\protect\rule[1.1ex]{.325em}{.1ex}}h%
\omega ,  \label{quantum}
\end{equation}
when the radiation involved has an angular frequency $\omega .$ A heuristic
derivation of this relation will be seen in section 5.2 below, as due the
interference between any given radiation and the vacuum field.

Our aim here is to explain the observed spectrum of the hydrogen atom, which
was the main success of Bohr atomic model. In order to get a physical
picture of the first Bohr postulate we will study just circular orbits. We
may assume that orbits relatively stable have discrete energies $E_{n},$ $n$
being an integer and $n=1$ corresponding to the ground state eq.$\left( \ref
{04}\right) .$ Actually the existence of a discrete set of (almost) stable
orbits cannot be easily derived from our assumption of a real vacuum
radiation, but if we accept this assumption the full spectrum of energies of
the atom may be derived as follows. We shall study transitions between two
close orbits taking eq.$\left( \ref{quantum}\right) $ into account, that is 
\begin{mathletters}
\begin{equation}
E_{n+1}-E_{n}=
\rlap{\protect\rule[1.1ex]{.325em}{.1ex}}h%
\omega _{n+1\rightarrow n},  \label{trans}
\end{equation}
where $\omega _{n+1\rightarrow n}$ is the frequency of the absorbed or
emitted radiation. Now it is plausible to assume that the radiation
frequency is related to the rotation frequency of the electron, whence we
may write 
\end{mathletters}
\begin{equation}
E_{n+1}-E_{n}=
\rlap{\protect\rule[1.1ex]{.325em}{.1ex}}h%
\omega _{n+1\rightarrow n}\simeq \frac{1}{2}\left( \omega _{n+1}+\omega
_{n}\right) 
\rlap{\protect\rule[1.1ex]{.325em}{.1ex}}h%
,  \label{trans1}
\end{equation}
where the $\omega _{n+1\rightarrow n}$ is the frequency of an emitted or
absorbed radiation and $\omega _{n}$ , $\omega _{n+1}$ are rotation
frequencies in the atomic states $n$ and $n+$1 respectively. The frequencies
may be related to the energies according to classical electrodynamics (see
the former three eqs.$\left( \ref{03}\right) )$, that is 
\[
\omega _{n}^{2}=\frac{8\left| E_{n}\right| ^{3}}{me^{4}}, 
\]
whence eq.$\left( \ref{trans1}\right) $ becomes 
\begin{equation}
\left| E_{n+1}\right| -\left| E_{n}\right| \simeq -\frac{\sqrt{2}\left(
\left| E_{n+1}\right| +\left| E_{n}\right| \right) ^{3/2}}{\sqrt{m}e^{2}},
\label{Bohr2}
\end{equation}
which provides the set of energies for the stable states in Bohr model. An
approximate solution of eq.$\left( \ref{Bohr2}\right) $ may be easily
obtained when $n>>1$. In fact, we may take the variable $n$ as continuous
and substitute the following differential equation for eq.$\left( \ref{Bohr2}%
\right) ,$ that is 
\begin{equation}
\frac{d\left| E_{n}\right| }{dn}=\frac{2\sqrt{2}\left| E_{n}\right| ^{3/2}}{%
\sqrt{m}e^{2}n}.  \label{Bohr3}
\end{equation}
The solution of the differential equation is 
\begin{equation}
\left| E_{n}\right| =\frac{me^{4}}{2
\rlap{\protect\rule[1.1ex]{.325em}{.1ex}}h%
^{2}n^{2}},  \label{Bohrenergies}
\end{equation}
where the integration constant is fixed so that the ground state energy $%
E_{1}$ agrees with eq.$\left( \ref{04}\right) .$ It may be seen that eq.$%
\left( \ref{Bohrenergies}\right) $ is also an approximate solution of eq.$%
\left( \ref{Bohr2}\right) $ and it is equivalent to Bohr\'{}s first
postulate which states that circular orbits with angular momentum $L$ are
stable if 
\begin{equation}
L=n
\rlap{\protect\rule[1.1ex]{.325em}{.1ex}}h%
.  \label{Bohr}
\end{equation}

Actually eqs.$\left( \ref{trans1}\right) $ to $\left( \ref{Bohr3}\right) $
are well known from long ago. Indeed the fact that the frequency of emitted
or absorbed radiation agrees with the rotational frequency of the electron
for Bohr orbits with large $n$ is a typical example of Bohr\'{}s
correspondence principle.

The conclusion of this section is that classical electrodynamics combined
with the assumption of a (vacuum) radiation field filling space suggests an
intuitive picture for the quantization of the hydrogen atom, which might be
generalized to other quantum systems.

\subsection{The Casimir effect}

The Casimir effect consists of the attraction between two parallel perfectly
conducting plates in vacuum. The force $F$ per unit area depends on the
distance $l$ between the plates, 
\begin{equation}
F=-\frac{\pi ^{2}
\rlap{\protect\rule[1.1ex]{.325em}{.1ex}}h%
c}{240l^{4}},  \label{Casimir}
\end{equation}
a force confirmed empirically\cite{Milonnibook}. The reason for the
attraction may be understood qualitatively as follows. In equilibrium the
electric field of the vacuum radiation (that we will label zeropoint field,
ZPF) should be nil on any plate surface, otherwise an electric current would
be produced. This fact constrains the possible normal modes of the
radiation, mainly those having wavelengths $\lambda \gtrsim l$, but the
distribution of high frequency (short wavelengths) modes would be barely
modified by the presence of the plates. If we assume that an effective
cut-off exists for $\lambda \geq Kl$ then the decrease of energy of the ZPF
in the space between plates becomes 
\[
E\sim A\times l\times \int_{0}^{2\pi c/Kl}\frac{%
\rlap{\protect\rule[1.1ex]{.325em}{.1ex}}h%
}{2\pi ^{2}c^{3}}\omega ^{3}d\omega =\frac{2\pi ^{2} 
\rlap{\protect\rule[1.1ex]{.325em}{.1ex}}h%
c}{K^{4}l^{3}}A, 
\]
where \textit{A} is the area of the plates and eq.$\left( \ref{01}\right) $
has been taken into account. The derivative of $E/A$ with respect to the
distance $l$ agrees with eq.$\left( \ref{Casimir}\right) $ if $K\simeq 6$.

The rigorous SED derivation \cite{Dice},\cite{Santos20}, \cite{book}\ is
similar to the quantum-mechanical calculation just substituting stochasic
averages for quantum vacuum expectations. It consists of determining the
normal modes of the radiation when the plates are at a distance $l$ and then
to attribute a mean enery $\frac{1}{2}
\rlap{\protect\rule[1.1ex]{.325em}{.1ex}}h%
\omega $ to every mode. The energy diverges if we sum over all radiation
modes but the force per unit area is finite and it reproduces eq.$\left( \ref
{Casimir}\right) $. A regularization procedure is required in order to get
the result\cite{Milonnibook}. The physical picture of the phenomenon is that
the radiation pressures in both faces of each plate are different and this
is the reason for a net force on the plate. The Casimir effect is currently
considered the most strong argument for the reality of the quantum vacuum
fields. For us it is specially relevant because it provides an example of
the fact that what matters is the difference between the radiation arriving
at the two faces of a plate, rather than the total radiation acting on one
side. A similar behaviour will be assumed for photocounters in section 7.4.

\section{The vacuum radiation as a stochastic field}

\subsection{The probability distribution of the field amplitudes}

The spectrum eq.$\left( \ref{01}\right) $ does not characterize completely
the background field, it is necessary to know relevant probability
distributions. A standard method to determine the properties of radiation is
to expand the field in normal modes, which may be defined in a fixed
normalization volume $V$, usually a cube with side $L$. Eventually we should
go to the limit $V\rightarrow \infty .$ A usual expansion for the
electromagnetic field starts with the vector potential, $\mathbf{A(r,}t%
\mathbf{)}$, that may be written, in the Coulomb gauge with rationalized
units, as follows 
\begin{equation}
\mathbf{A}=\frac{1}{\sqrt{V}}\sum_{j}\sqrt{\frac{%
\rlap{\protect\rule[1.1ex]{.325em}{.1ex}}h%
c^{2}}{2\omega _{j}}}\left[ a_{j}\mathbf{\varepsilon }_{j}\exp \left( i%
\mathbf{k}_{j}\cdot \mathbf{x-}i\omega _{j}t\right) +a_{j}^{*}\mathbf{%
\varepsilon }_{j}\exp \left( -i\mathbf{k}_{j}\cdot \mathbf{x+}i\omega
_{j}t\right) \right] ,  \label{f5}
\end{equation}
where $\mathbf{k}_{j}$ is the wavevector, $\omega _{j}$ the frequency and $%
\mathbf{\varepsilon }_{j}$ the (linear) polarization vector of the normal
mode labeled by \textit{j}. The dimensionless complex quantities $a_{j}$ and 
$a_{j}^{*}$ are named the amplitudes of the mode. Hence the electric, $%
\mathbf{E}$, and magnetic, $\mathbf{B}$, fields of the radiation may be
obtained via 
\begin{equation}
\mathbf{E=-}\frac{1}{c}\frac{\partial \mathbf{A}}{\partial t},\mathbf{%
B=\bigtriangledown \times A.}  \label{f6}
\end{equation}
The radiation energy in the volume $V$ becomes 
\begin{equation}
H=\frac{1}{2}\int \left( \mathbf{E\cdot E+B\cdot B}\right)
d^{3}r=\sum_{j}\left( 
\rlap{\protect\rule[1.1ex]{.325em}{.1ex}}h%
\omega _{j}\left| a_{j}\right| ^{2}\right) ,  \label{f1}
\end{equation}
whose asociated mean energy fits in eq.$\left( \ref{02}\right) ,$ provided
that the expectation value for every mode fulfils 
\begin{equation}
\left\langle \left| a_{j}\right| ^{2}\right\rangle =1/2.  \label{f0}
\end{equation}
Of course the number of normal modes diverges so that some regularization
would be requiered for high frequencies, a problem also known in quantum
field theory which we will comment on section 6.2.

The stochastic field is fully characterized by the joint probability
distribution of the mode amplitudes, $\rho \left( \left\{
a_{j},a_{j}^{*}\right\} \right) $, a distribution that should be compatible
with eqs.$\left( \ref{f0}\right) $ and $\left( \ref{01}\right) .$ These
equations suggest that different modes are statistically independent whence
we should write $\rho \left( \left\{ a_{j},a_{j}^{*}\right\} \right) $ as a
product of probabilities of the modes. Finally it is plausible to assume
that the probability of every mode amplitude is Gaussian with random phases,
that is the phase of $a_{j}$ is distributed uniformly in $\left( 0,2\pi
\right) .$ These constraints lead to the following normalized joint
probability distribution

\begin{equation}
\rho \left( \left\{ a_{j},a_{j}^{*}\right\} \right) =\prod_{j}\frac{2}{\pi }%
\exp \left( -2\left| a_{j}\right| ^{2}\right) .  \label{f3}
\end{equation}
Hence we may get the expectation of any function of the vacuum electric and
magnetic fields which, taking eqs.$\left( \ref{f5}\right) $ and $\left( \ref
{f6}\right) $ into account, can be reduced to an integral of a function $%
f\left( \left\{ a_{j},a_{j}^{*}\right\} \right) $ of the amplitudes weighted
by the distribution eq.$\left( \ref{f3}\right) .$ That is 
\begin{equation}
\left\langle f\left( \left\{ a_{j},a_{j}^{*}\right\} \right) \right\rangle
=\int f\left( \left\{ a_{j},a_{j}^{*}\right\} \right) \rho \left( \left\{
a_{j},a_{j}^{*}\right\} \right) \prod_{j}d\func{Re}a_{j}d\func{Im}a_{j}.
\label{f7}
\end{equation}
As an example let us consider the function of a single mode $%
f=a_{j}^{n}a_{j}^{*m}$. The integral is different from zero only if $m=n$
due to the random phases and we get 
\begin{equation}
\left\langle a_{j}^{n}a_{j}^{*m}\right\rangle =\delta _{nm}\int_{0}^{\infty
}\left| a_{j}\right| ^{2n}\frac{2}{\pi }\exp \left( -2\left| a_{j}\right|
^{2}\right) \pi d\left| a_{j}\right| ^{2}=2^{-n}n!\delta _{nm},  \label{f8}
\end{equation}
$\delta _{nm}$ being Kronecker\'{}s delta.

\subsection{Other states of the field}

Now let us study modifications of the vacuum radiation defined by the
probability distribution eq.$\left( \ref{f3}\right) $. The modified states
would have a different probability distribution than the vacuum ones and
they could correspond to what in quantum language are called excited states.
The modification may consist of the addition of some radiation to the vacuum
state. In particular if the added field has a probability distribution of
modes $\rho _{1}\left( \left\{ a_{j},a_{j}^{*}\right\} \right) ,$
uncorrelated with the vacuum $\rho \left( \left\{ a_{j},a_{j}^{*}\right\}
\right) ,$ then the total radiation will have a probability distribution
that is the convolution of the distributions $\rho $ and $\rho _{1}.$ In the
following we study a simple example in order to show that the addition of
radiation to the vacuum may modify dramatically the field because we should
add the amplitudes rather than the intensities$.$

Let us consider a single vacuum mode with amplitude $a_{j\text{ }}$and the
addition of another sure (i.e. not random) amplitude $b_{j}.$ The
distribution of the total mode amplitude, 
\begin{equation}
c_{j}=a_{j}+b_{j},  \label{9f}
\end{equation}
will be 
\begin{eqnarray}
\rho _{c}\left( c_{j},c_{j}^{*}\right) &=&\frac{2}{\pi }\exp \left( -2\left|
a_{j}\right| ^{2}\right) =\frac{2}{\pi }\exp \left( -2\left|
c_{j}-b_{j}\right| ^{2}\right)  \nonumber \\
&=&\frac{2}{\pi }\exp \left( -2\left| c_{j}\right| ^{2}\right) \times \exp
\left[ 4\func{Re}(c_{j}b_{j}^{*})-2\left| b_{j}\right| ^{2}\right] .
\label{f9}
\end{eqnarray}
It may be realized that the expectation of the intensity in the mode
increases by the intensity of the field added, that is 
\[
\left\langle \left| c_{j}\right| ^{2}\right\rangle =\frac{1}{2}+\left|
b_{j}\right| ^{2}. 
\]
The phases are also modified, i.e. they are no longer uniformly distributed.

Another interesting example is the state produced by a change of the
distribution of phases in one or several modes. For instance we may replace
the uniform phase distribution in one of the modes, this giving a squeezed
vacuum state of light. If in addition to the phase change the intensity is
also increased, we would have a usual squeezed state. We will not study
these possibilities further on.

I shall point out that not all density operators (or state vectors) that are
assumed to represent states of the radiation in the quantum (Hilbert space)
formalism may correspond to states in our stochastic interpretation. An
obvious requirement is that the joint density function of mode amplitudes is
positive in order that it may be interpreted as a probability distribution.
I propose that quantum ``states'' not fulfilling this condition are not
physical states. In particular this is the case for the socalled photon
number states (or Fock states) consisting of a fixed number of ``photons''.
I stress again that in the approach supported in this article photons are
just mathematical objects useful for calculations, not physical states in
general. I believe that the demand that all quantum states (e.g. those with
an integer number of photons) are physical states makes impossible to get an
intuitive picture of the (quantum) radiation field.

\section{ The particle behaviour of light}

In this section we shall show that the vacuum radiation, taken as a
stochastic field, provides hints for a realistic interpretation of the
particle behaviour of light.

\subsection{What is a photon?}

Maxwell theory establishes that light consists of electromagnetic waves.
However this view was allegedly superseded by the proposal that light
consists also of particles, later named photons. The wave-particle behaviour
is the main mystery of quantum mechanics, in the words of Feynman, and it
prevents a clear understanding of the theory. In the following I provide
qualitative explanations for some examples of particle behaviour of light
within Maxwell theory. That behaviour may be understood as due to the
existence of the vacuum stochastic radiation studied in the previous section.

In the year 1900 Planck assumed that energy exchanges between matter and
light take place in discrete amounts (``quanta'') of energy related to the
frequency by 
\begin{equation}
E=
\rlap{\protect\rule[1.1ex]{.325em}{.1ex}}h%
\omega .  \label{Planck}
\end{equation}
Five years later Einstein went further postulating that light itself
consists of particles with energy $
\rlap{\protect\rule[1.1ex]{.325em}{.1ex}}h%
\omega $, whence he derived the law of the photoelectric effect. Actually
Planck assumption eq.$\left( \ref{Planck}\right) $ is sufficient to derive
that law, without the stronger Einstein postulate. In fact assuming that
when monocromatic light with frequency $\omega $ arrives at an appropriate
material only a quanta $
\rlap{\protect\rule[1.1ex]{.325em}{.1ex}}h%
\omega $ may be absorbed at a time, a part $E_{0}$ used to extract an
electron and the rest to supply it kinetic energy $E_{k}$. Then we have 
\[
E_{k}=
\rlap{\protect\rule[1.1ex]{.325em}{.1ex}}h%
\omega -E_{0}\text{ if }
\rlap{\protect\rule[1.1ex]{.325em}{.1ex}}h%
\omega >E_{0},\text{ no effect otherwise,} 
\]
which is the law of the photoelectric effect. The constraint that radiation
may be absorbed only in amounts fulfilling eq.$\left( \ref{Planck}\right) $
is the first example of particle-like behaviour, that we will explain
qualitatively in section 5.2.

In the celebrated 1916 article about the absorption and emission of
radiation, Einstein arrived at the conclusion that the radiation emitted by
an atom possesses well defined momentum, in his words it appears in the form
of radiation needles. The two commented claims by Einstein led to the
popular belief that light consists of particles (photons) with definite
energy and momentum each. Compton experiments in 1923-24 are commonly viewed
as a confirmation of that belief. A semi-quantitative explanation of the
radiation needles will be provided in section 5.3.

More recently experiments have been performed in optics that dramatically
exhibit wave-particle behaviour of light. I will comment on them in section
5.4.

\subsection{Understanding quanta: Discrete energy exchanges}

Firstly I point out that the absorption of light in the form of localized
spots in a photographic plate or clicks in a photodetector are not valid
arguments for the particle behaviour of radiation. In fact the former is
caused by the granular (atomic or molecular) nature of photografic plates.
The latter derive from the fact that photocounters are manufactured so that
they click when the radiation arriving during a detection time surpasses
some threshold, which is compatible with light being continuous (waves). For
a model of detector see section 7.4

The absorption of light in discrete amounts, fulfilling eq.$\left( \ref
{Planck}\right) ,$ may be understood as follows. Let us consider a light
signal with wavevector $\mathbf{k}_{0}$ that arrives at a material having
weakly bound electrons. The vacuum radiation may be described in terms of
plane waves as in eq.$\left( \ref{f5}\right) $ and we are interested in
those waves with wavevectors $\mathbf{k}$ near $\mathbf{k}_{0}$. From time
to time it may happen that several of these waves have phases close to the
incoming signal, whence they will interfere constructively giving rise to a
unusually large intensity during some time $T$. In this case a transfer of
energy to the material will be more probable, for example an electron may be
ejected. We may identify $T$ with the coherence time of those radiation
consisting of the signal plus the vacuum radiation able to interfere
constructively with it. The question is how much energy $E$ may be
transferred.

The wavevectors $\mathbf{k}$ of the vacuum field effective for the transfer
of energy should be close to $\mathbf{k}_{0}$ in order that interference
takes place. That is $\left| \mathbf{k-k}_{0}\right| <<\left| \mathbf{k}%
_{0}\right| \equiv k_{0}$ whence $\mathbf{k}$ may differ but slightly from $%
\mathbf{k}_{0},$ either in modulus or direction or both. Thus we may write 
\[
\left| k-k_{0}\right| =\delta k\sim c\delta \omega <<c\omega ,(\mathbf{k-k}%
_{0})_{\perp }/k=\sin \theta \simeq \theta <<1, 
\]
where $(\mathbf{k-k}_{0})_{\perp }$ means the component of $\mathbf{k-k}_{0}$
perpependicular to $\mathbf{k.}$ It is plausible to identify $\omega
=k_{0}/c,\theta \simeq \delta \omega /\omega $ and $T\simeq \pi /\delta
\omega .$ We need the area $A$ effective for absorption, that may be
estimated as follows 
\[
A\sim \frac{1}{2}L^{2}\theta ^{2}=\frac{1}{2}T^{2}c^{2}\delta \omega
^{2}/\omega ^{2}=\frac{1}{2}\left( \frac{\pi c}{\omega }\right) ^{2}, 
\]
where $L=Tc$ is the coherence length. The effective intensity (energy per
unit area per unit time) should be 
\[
I=cS\left( \omega \right) \Delta \omega , 
\]
where $S\left( \omega \right) $ is the spectrum (energy per unit volume and
unit frequency interval). Thus we get for the absorbed energy

\begin{equation}
E\sim ITA\simeq cS\Delta \omega \times \left( \pi /\Delta \omega \right)
\times \frac{1}{2}(\pi c/\omega )^{2}\simeq S\frac{\pi ^{3}c^{3}}{2\omega
^{2}}.  \label{p1}
\end{equation}
$S\left( \omega \right) $ may be of order the vacuum field spectrum given in
eq.$\left( \ref{01}\right) $, which leads to 
\begin{equation}
E\sim \frac{
\rlap{\protect\rule[1.1ex]{.325em}{.1ex}}h%
}{2\pi ^{2}c^{3}}\omega ^{3}\times \frac{\pi ^{3}c^{3}}{2\omega ^{2}}=\frac{%
\pi }{4}
\rlap{\protect\rule[1.1ex]{.325em}{.1ex}}h%
\omega ,  \label{p2}
\end{equation}
for the energy that may be transferred to the electron, in rough agreement
with eq.$\left( \ref{Planck}\right) .$

\subsection{Radiation needles and the Compton effect}

An interpretation of the needle radiation that appears in the emission of
light by atoms is as follows. In our stochastic interpretation the emission
is not spontaneous but induced by the vacuum field (or zeropoint field,
ZPF). Then let us assume that in a fluctuation a strong plane wave of the
ZPF with frequency $\omega $ arrives at an atom and it happens that $\omega $
is also one of the possible frequencies for emission from the excited atom.
Then the arriving plane wave component of the ZPF may induce the emission of
radiation with the same frequency and phase than the incoming wave. Thus the
emitted radiation should correspond to the addition of the amplitudes (not
the intensities!) of the incoming plane wave plus the emitted spherical
wave. The frequencies being equal there would be interference and it is not
difficult to show that it will be constructive in the forward direction and
mainly destructive in all other directions.

More quantitatively, the outgoing energy will be concentrated within the
region where the phase difference is small. The boundary of that region is
roughly defined by the following relation with the half angle, $\theta $, as
seen from the atom and the distance, $d$, that is 
\begin{equation}
\frac{d}{\cos \theta }-d\sim \frac{\lambda }{2}\Rightarrow \theta \sim \sqrt{%
\frac{\lambda }{d}},  \label{needle}
\end{equation}
where $\lambda $ is the wavelength. If we take $d$ to be coherence length of
the emitted ``photon'', for typical atomic emissions we have $d\sim 1$m, $%
\lambda \sim 1\mu ,$ so that $\theta \sim 10^{-3}$. This fits with
Einstein\'{}s proposal of ``needles of radiation'' and, in addition, it
explains the random character of the direction of emission. In our
interpretation the stochastic character of the ZPF is the cause of the
randomness.

This provides the picture of a localized photon as a concentration of
radiation energy that nevertheless has a frequency relatively well defined.
Furthermore that frequency is plausibly related to the emitted energy by
Planck eq.$\left( \ref{Planck}\right) ,$ taking the arguments of the
previous section into account. In any case the coherence time of the
radiation needle cannot be larger than the lifetime of the atomic state.

We may apply that photon model to the case of an atomic cascade where two
photons are emitted within a short time interval. Then the picture that
emerges is the existence of two ``needles of radiation'' moving in different
directions. In particular if both the initial and the final state of the
atom have zero spin and the photons are emitted in opposite directions, then
the angular momenta of the two photons should be opposite by angular
momentum conservation, whence they will be strongly correlated in
polarization. The quantum formalism predicts that they will be maximally
entangled, but I will not provide an interpretation of photon entanglement
at this moment, see below section 7.5. The polarization correlation will
diminish if the photons are emitted at an angle smaller than 180${{}^{o}}$,
and this causes that no Bell inequality may be violated in experiments using
photon pairs from atomic cascades\cite{Santos1992}. Several atomic cascade
tests of the Bell inequalities were performed in the decade 1975-1985.

As another example I propose a semi-quantitative model for the Compton
effect. As is well known Compton\'{}s was the experiment that the scientific
community accepted as the final proof of the existence of photons. The
experiment is usually understood as a collision between one photon of X-ray,
frequency $\omega _{1},$ and one electron, giving rise to another photon
with smaller frequency, $\omega _{2},$ at an angle $\theta $ with the
incident one and a recoil electron. Indeed the (relativistic) kinematics may
be explained assuming that the incident and outgoing photons have energies $
\rlap{\protect\rule[1.1ex]{.325em}{.1ex}}h%
\omega _{1}$ and $
\rlap{\protect\rule[1.1ex]{.325em}{.1ex}}h%
\omega _{2},$ respectively, and the electron is initially at rest. Quantum
electrodynamics gives a quantitative account of the phenomenon, including
the cross section of the process, but it does not offer an intuitive
picture. On the other hand there have been several attempts at a
semiclassical explanation that I will not revisit here.

A stochastic interpretation might be achieved if we substitute radiation
needles for photons. A rough model is as follows. Let us consider an
incoming monocromatic X-ray beam with frequency $\omega _{1}$. By the
arguments leading to eqs.$\left( \ref{p2}\right) $ we may assume that the
beam contains radiation wavepackets with energy $
\rlap{\protect\rule[1.1ex]{.325em}{.1ex}}h%
\omega _{1}$ and momentum $
\rlap{\protect\rule[1.1ex]{.325em}{.1ex}}h%
\omega _{1}/c$. From time to time a large fluctuation of the ZPF may cross
at an angle $\theta $ the incoming X-ray beam in a region where there are
weakly bound electrons. If the ZPF fluctuation has an appropriate frequency
it could interfere with the radiation of the X-ray beam producing a
concentration of energy in a direction at an angle $\theta _{1}$\TEXTsymbol{<%
}$\theta ,$ which may accelerate one electron in that direction$.$ The
electron will radiate with energy and momentum determined by the
conservation laws.

\subsection{The wave-particle behaviour of light in optics}

In quantum optics the experiments may be usually interpreted in terms of
light waves, the particle behaviour been apparent only in photodetection.
Detectors will not be studied here in detail (see section 7.4) but we may
plausibly assume that the particle behaviour of light in detection is
usually related to the corpuscular nature of atoms, or electrons in
detectors. However there are cases when this explanation is not sufficient
or not approprate. We will study two examples: anticorrelation after a
beam-splitter in the following and entangled photon pairs in section 7.4.

A simple beam-splitter (BS) may just consist of a slab of transparent
material. If a light beam impinges at a point of the slab, a part of the
beam intensity is transmitted and another part reflected. The relative
intensities of the outgoing fields depend on the refraction index of the
material and the angle of incidence. In this way we have an elementary
beam-splitter with one incoming channel and two outgoing channels. Actually
we have another incoming channel via a light beam arriving in the opposite
side of the slab that, gives rise to two new outgoing channels. In practice
the plate is used so that the transmitted light from the first incoming
channel is superposed to the reflected light of the second incoming channel
and the light reflected from the former is superposed to the light
transmitted from the latter. In this way we would have two incoming channels
and two outgoing ones. In practice beam-splitters may be more sophisticated,
e. g. involving piles of plates (used for instance in many tests of Bell
inequalities). Sometimes the BS\ polarizes the light, thus acting as a
polarizer or a polarization analyzer.

In the following I study in more detail \textit{a balanced non-polarizing BS}%
. If the field amplitudes of the incoming beams are $E_{1}$ and $E_{2}$,
then the amplitudes in the outgoing channels will be 
\begin{equation}
E_{out1}=\frac{1}{\sqrt{2}}\left( E_{1}+iE_{2}\right) ,E_{out2}=\frac{1}{%
\sqrt{2}}\left( E_{2}+iE_{1}\right) .  \label{BS}
\end{equation}
The imaginary unit $i$ is appropriate if we treat the electromagnetic fields
in the complex representation, as we will made throughout this article. From
eq.$\left( \ref{BS}\right) $ it is obvious that the energy is conserved in
the BS. In fact the sum of intensities in the incoming channels equals the
similar sum in the outgoing ones, that is 
\[
I_{out1}+I_{out2}=\left| E_{out1}\right| ^{2}+\left| E_{out2}\right|
^{2}=\left| E_{1}\right| ^{2}+\left| E_{2}\right| ^{2}=I_{in1}+I_{in2}. 
\]
In the experiment studied in the following the field arriving at one of the
incoming channels will be a signal and a vacuum field at the other one.

\subsection{Anticorrelation-recombination experiment}

A dramatic exhibition of the wave-particle behaviour of light is the
anticorrelation-recombination experiment \cite{Grangier}. A weak radiation
signal, allegedly consisting of well separated photons, is sent to one of
the incoming channels of a balanced beam splitter $BS1,$ and two
photodetectors, $A$ and $B$, are placed in front of the outgoing channels.
No coincidences are observed, which allegedly shows the corpuscle behaviour
of light: a photon is not divided, but goes to one of the detectors. If the
detectors are removed and the two outgoing radiation beams are recombined
via the two incoming channels of another beam splitter $BS2$, then the
detection in one of the outgoing channels depends on the length difference
between the two paths from $BS1$ to $BS2$, this being a typical wave
behaviour.

Our stochastic intepretation is as follows\cite{MS1}. If we assumed that the
vacuum quantum fields were not real fields then only the signal field $E$
entering $BS1$ should produce outgoing fields, in every one of the two
outgoing channels$.$ However if the vacuum fields are real, there is another
(vacuum) field $E_{0}$ with similar frequency than the signal entering in $%
BS1$ via the second incoming channel and interference is produced. Hence the
outgoing fields may be written 
\begin{equation}
E_{A}=\frac{E+iE_{0}}{\sqrt{2}},E_{B}=\frac{iE+E_{0}}{\sqrt{2}}.  \label{10}
\end{equation}
Depending on the relative phases, one of the intensities may be large and
the other one small, that is 
\begin{eqnarray}
I_{A} &=&\left| E_{A}\right| ^{2}=\frac{1}{2}(\left| E\right| ^{2}+\left|
E_{0}\right| ^{2})+\left| E\right| \left| E_{0}\right| \cos \phi ,
\label{m9} \\
I_{B} &=&\left| E_{B}\right| ^{2}=\frac{1}{2}(\left| E\right| ^{2}+\left|
E_{0}\right| ^{2})-\left| E\right| \left| E_{0}\right| \cos \phi ,  \nonumber
\end{eqnarray}
where $\phi $ is the relative phase of the fields $E$ and $E_{0}.$ On the
other hand the vacuum intensities would be ideally $I_{A0}=$ $%
I_{B0}=I_{0}=\left| E_{0}\right| ^{2}$.

If we assume that detection is roughly proportional to the part of the
arriving intensity that surpasses the ZPF level, then with $I=\left|
E\right| ^{2},I_{0}=\left| E_{0}\right| ^{2},$ we obtain 
\[
R_{A}=\frac{1}{2}\left\langle I-I_{0}\right\rangle ,R_{AB}=\frac{1}{4}%
\left\langle (I-I_{0})^{2}\right\rangle -\frac{1}{2}\left\langle
I\right\rangle \left\langle I_{0}\right\rangle . 
\]
This result shows that for weak signals, that is when $I$ is not much
greater than $I_{0},$ the coincidence detection rate is inhibited, that is $%
R_{AB}<<R_{A}R_{B},$ as observed in the commented experiment. (We define the
rate as a dimensionless probability of detection per time window). In
contrast for macroscopic (classical) light we have $I>>I_{0}$ and the ratio
would be 
\[
r\equiv \frac{R_{AB}}{R_{A}R_{B}}\simeq \frac{\left\langle
I^{2}\right\rangle }{\left\langle I\right\rangle ^{2}}. 
\]
Hence if the radiation has fixed (nonfluctuating) intensity, like the laser
light, then 
\[
\left\langle I^{2}\right\rangle =\left\langle I\right\rangle ^{2}\Rightarrow
r=1 
\]
meaning that the detections are uncorrelated. On the other hand for chaotic
light, where the field fluctuations are Gaussian, we would have 
\[
\left\langle I^{2}\right\rangle =2\left\langle I\right\rangle
^{2}\Rightarrow r=2, 
\]
meaning that the detections by Alice and Bob are positively correlated. The
change from $r=1$ to $r=2,$ a phenomenon known as ``photon bunching'', has
been interpreted as a quantum effect attributed to the Bose character of
photons. In our stochastic interpretation it is the consequence of the
correlated fluctuations derived from the Gaussian character of chaotic light.

In the recombination process the fields eq.$\left( \ref{10}\right) $ will
enter $BS2$ giving rise to the following intensity in one of the outgoing
channels 
\begin{equation}
I_{rec}=\left| \frac{\left| E_{A}\right| }{\sqrt{2}}+\exp \left( i\theta
\right) \frac{\left| E_{B}\right| }{\sqrt{2}}\right| ^{2}=\frac{1}{2}\left(
I+I_{0}\right) +\frac{1}{2}\left( I-I_{0}\right) \cos \theta ,  \label{m13}
\end{equation}
where $\theta $ is the relative phase due to the different path lengths. The
device used in the experiment\cite{Grangier}, consisting of two beam
splitter and two mirrors in between, is called Mach-Zehnder interferometer.
The detection rate is proportional to 
\[
R_{A}=\left\langle I_{rec}\right\rangle -\left\langle I_{0}\right\rangle =%
\frac{1}{2}\left\langle I-I_{0}\right\rangle \left( 1+\cos \theta \right) , 
\]
meaning that a 100\% visibility may be achieved. Thus we have a wave
explanation for one of the most dramatic particle behaviour of light, the
anticorrelation after a beam splitter. The anticorrelation is usually named
``photon antibunching'' and it is considered a typically quantum phenomenon,
that cannot be explained by classical theories. Of course it can be
explained if we do assume that the vacuum fields are real stochastic fields.
The evolution of these fields is classical (Maxwellian) but the assumption
of real vacuum fields is alien to classical physics. I stress that Planck
constant appears fixing the scale of the vacuum fields, see eq.$\left( \ref
{f3}\right) .$

\section{Connection with the quantum formalism}

\subsection{The vacuum in the Hilbert space formalism}

The existence of radiation in vacuum, even at zero Kelvin, appeared for the
first time in an extension by W. Nernst of Planck's assumption about the
finite energy of oscillators in his second radiation theory of 1912. The
zeropoint energy of the electromagnetic field (ZPF) was disregarded because
it is divergent, although the consequences of its possible reality were
initially explored by several authors, including Einstein and Nernst\cite
{Milonnibook}. Soon sfterwards the ZPF was forgotten due to the advent of
Bohr atomic model in 1913, that opened a new route for the ``old quantum
theory'' which was followed by the mainstream of the community.

The ZPF reappeared in 1927 when Dirac quantized the electromagnetic field
staarting from an expansion in normal modes, see eq.$\left( \ref{f5}\right)
, $ then promoting the amplitudes to be operators $\left\{ \hat{a}_{j},\hat{a%
}_{j}^{\dagger }\right\} $ in a Hilbert space. These operators are usually
named ``annihilation and creation operators of photons'' in the mode, and
fulfil the commutation rules 
\begin{equation}
\hat{a}_{j}\hat{a}_{k}=\hat{a}_{k}\hat{a}_{j},\hat{a}_{j}^{\dagger }\hat{a}%
_{k}^{\dagger }=\hat{a}_{k}^{\dagger }\hat{a}_{j}^{\dagger },\hat{a}_{j}\hat{%
a}_{k}^{\dagger }=\hat{a}_{k}^{\dagger }\hat{a}_{j}+\delta _{jk},  \label{h3}
\end{equation}
where $\delta _{jk}$ is the Kronecker delta. In the formalism the
Hamiltonian operator of the field may be written as a sum over normal modes,
that is 
\begin{equation}
H=\frac{1}{2}
\rlap{\protect\rule[1.1ex]{.325em}{.1ex}}h%
\sum_{j}\omega _{j}\left( \hat{a}_{j}+\hat{a}_{j}^{\dagger }\right) ^{2}=%
\frac{1}{2}
\rlap{\protect\rule[1.1ex]{.325em}{.1ex}}h%
\sum_{j}\omega _{j}(\hat{a}_{j}^{2}+\hat{a}_{j}^{\dagger 2}+\hat{a}%
_{j}^{\dagger }\hat{a}_{j}+\hat{a}_{j}\hat{a}_{j}^{\dagger }),  \label{0}
\end{equation}
where $\omega _{j}$ is the (angular) frequency of the normal mode $j$. The
energy is given by the vacuum expectation of the Hamiltonian, that is 
\begin{eqnarray}
\left\langle 0\left| H\right| 0\right\rangle &=&\frac{1}{2} 
\rlap{\protect\rule[1.1ex]{.325em}{.1ex}}h%
\sum_{j}\omega _{j}\left\langle 0\left| (\hat{a}_{j}^{2}+\hat{a}%
_{j}^{\dagger 2}+2\hat{a}_{j}^{\dagger }\hat{a}_{j}\right| 0\right\rangle +%
\frac{1}{2}
\rlap{\protect\rule[1.1ex]{.325em}{.1ex}}h%
\sum_{j}\omega _{j}\left\langle 0\left| 1\right| 0\right\rangle  \nonumber \\
&=&\frac{1}{2}
\rlap{\protect\rule[1.1ex]{.325em}{.1ex}}h%
\sum_{j}\omega _{j},\smallskip  \label{a}
\end{eqnarray}
the former expectation being nil. The result corresponds to a mean energy $%
\frac{1}{2}
\rlap{\protect\rule[1.1ex]{.325em}{.1ex}}h%
\omega _{j}$ per mode, that fits in the arguments of section 3.2 above.

Around 1947 two discoveries reinforced the hypothesis of the quantum vacuum
fields, namely the Casimir effect and the Lamb shift. The former has been
discussed in section 3.4. Lamb and Retherford observed an unexpected
absorption of microwave radiation by atomic hydrogen, that was soon
explained in terms of the interaction of the atom with the quantized
electromagnetic field, which involves the vacuum radiation (ZPF). Indeed
Willis Lamb has claimed to be the discoverer of the ZPF by experiment\cite
{Lamb1}. The finding led, in a few years, to the development of quantum
electrodynamics (QED), a theory that allows predictions in spectacular
agreement with experiments, and it was the starting point for the whole
theory of relativistic quantum fields. The success of QED rests on
renormalization techniques where it is taken for granted that particles,
like electrons, are dressed with ``virtual fields'' making their physical
mass and charge different from the bare quantities. In my view the
assumptions behind renormalization are actually a reinforcement of the
reality of the quantum vacuum fields, although people avoid commitement with
that conclusion using the word ``virtual'' as an alternative to ``really
existing''.

\subsection{The problem of the vacuum energy divergence}

The main problem with the ZPF is that the total energy density in space
diverges when we sum over all (infinitely many) modes. The standard solution
to the difficulty is to write the annihilation operators to the right. For
instance in eq.$\left( \ref{0}\right) $ to substitute $2\hat{a}_{j}^{\dagger
}\hat{a}_{j}$ for $\hat{a}_{j}^{\dagger }\hat{a}_{j}+\hat{a}_{j}\hat{a}%
_{j}^{\dagger }$. Then the vacuum expectation of the Hamiltonian would not
be eq.$\left( \ref{a}\right) $ but only its first term, which gives 0. Thus
the normal ordering is equivalent to choosing the zero of energies at the
level of the vacuum. It provides a practical procedure useful in
quantum-mechanical calculations, but for many authors it is not a good
solution. They see it as an ``ad hoc'' assumption, just aimed at removing
unpleasant divergences. For those authors the ZPF is a logical consequence
of quantization and the solution to the divergence problem should come from
a more natural mechanism.

In laboratory physics, where gravity usually plays no role, the possible
divergence of the quantum vacuum energy is not too relevant a question. In
fact their possibly huge, or divergent, energy may be usually ignored
choosing the zero energy at the vacuum level as said above. However this
choice is no longer innocuous in the presence of gravity because, according
to relativity theory, energy gravitates whence a huge vacuum energy should
produce a huge gravitational field. Therefore the possible existence of a
vacuum energy is a relevant question in astrophysics and cosmology.

A solution is to assume a cancelation between positive and negative terms in
the vacuum energy. Indeed it is the case that the vacuum electromagnetic
field, viewed as a stochastic field, contributes a positive energy and we
may extend this assertion to all vacuum contributions of Bose fields. In
fact we may assume that these fields can be also expanded in normal modes
and the vacuum should consists of a probability distribution of the
amplitudes similar to eq.$\left( \ref{f3}\right) .$ The vacuum contribution
of Fermi fields is quite different. We do not have a clear stochastic
interpretation, but there are arguments suggesting that they contribute
negative energy. The main reason is that the operators representing
amplitudes of the field obey anticommutation, rather than conmmutation,
rules. Thus it is plausible to assume that the Bose positive energy of the
vacuum is cancelled by the Fermi negative energy. In this article I will not
discuss the problem further on, but point out that the vacuum fields might
be the clue for understanding relevant problems in astrophysics, like the
nature of dark energy\cite{Santos11} or dark matter\cite{Santos18} and the
collapse of stellar objects\cite{Santos12}. A summary appears in chapter 7
of \cite{book}.

\subsection{Weyl transform and Wigner function in quantum mechanics}

After one century of successes we know that the Hilbert space formalism of
quantum mechanics (HS in the following) is extremely efficient in order to
deal with the microworld. As the main assumption in this article is that the
quantum vacuum fields are real stochastic fields, I believe that the HS
formalism could be understood as a disguised treatment of some peculiar
random variables. Consequently there should be a formalism alternative to HS
where the interpretation in terms of random variables is more clear. A
formalism exists that might do the job. It goes back to a proposal by
Hermann Weyl\cite{Weyl} in 1928, known as Weyl transform. Before discussing
the use of the transform for the study of the vacuum radiation, I will make
a digression about the application to mechanics of particles.

In fact Weyl proposed his transform for systems of particles with the design
of getting equations involving quantum operators, $\hat{x}_{j}$ and $\hat{p}%
_{j},$ from classical equations involving positions, $x_{j}$, and momenta, $%
p_{j}$, of particles. That is, his purpose was a quantization procedure of
nonrelativistic classical mechanics\cite{Scully}, \cite{Zachos}. Our aim in
this article is the opposite, namely to get classical-like probabilistic
equations from the quantum equations in order to get a realistic
interpretation, i.e. a picture of reality. Therefore we are interested in
the inverse Weyl transform that reads 
\begin{eqnarray}
f\left( \left\{ x_{j},p_{j}\right\} \right) &=&\frac{1}{\pi ^{2n}}%
\prod_{j=1}^{n}\int_{-\infty }^{\infty }d\lambda _{j}\int_{-\infty }^{\infty
}d\mu _{j}\exp \left[ -i(\lambda _{j}x_{j}+\mu _{j}p_{j})\right]  \nonumber
\\
&&\times Tr\left\{ \hat{f}\exp \left[ (i(\lambda _{j}\hat{x}_{j}+\mu _{j}%
\hat{p}_{j})\right] \right\} .  \label{W}
\end{eqnarray}
It provides a function $f\left( \left\{ x_{j},p_{j}\right\} \right) $ in
phase space for any trace class operator $\hat{f}$ in the Hilbert space.
Actually E. P. Wigner proposed in 1932 a transform equivalent to eq.$\left( 
\ref{W}\right) ,$ which gives rise to a new formalism for quantum mechanics
named Wigner representation \cite{Wigner}. In particular when $\hat{f}$ is
the density operator representing a quantum state, then eq.$\left( \ref{W}%
\right) $ gives the ``Wigner function'' of the state in the form of a
(pseudo-probability) distribution, \cite{Scully}, \cite{Zachos}.

The relevant question is whether eq.$\left( \ref{W}\right) $ provides a
realistic interpretation of the quantum states, and the answer seems to be
negative. In fact the Wigner function is not positive in general, whence it
cannot be interpreted as a probability distribution. We might assume that
quantum density operators represent physical states only when its Wigner
function is non-negative definite but this is too strong a restriction.
Actually the Wigner function cannot be interpreted as a phase space
distribution, in spite of it being a function of positions and momenta.
Indeed physical quantum particles cannot be just particles if we assume that
the vacuum fields are real. For instance a physical electron is not a small
(or pointlike) particle but a more complex system consisting of a particle
plus the modified vacuum fields interacting with it. Thus we might measure
the position of the particle, at least with some uncertainty whence the
probability distribution of positions $\rho \left( \mathbf{x}\right) $ has a
meaning and we may assume that it is obtained via the marginal of the Wigner
function, that is 
\[
\rho \left( \mathbf{x}\right) =\int W\left( \mathbf{x,p}\right) d^{3}\mathbf{%
p\geq }0. 
\]
Indeed $\rho \left( \mathbf{x}\right) $ agrees with the standard quantum
prediction for any quantum state of the particle. However the instantaneous
velocity would be highly irregular due to the interaction with the vacuum
fields whence the instantaneous momentum of the particle is meaningles, or
at least not measurable. We might determine the mean velocity (or momentum)
during some not too small time interval. Thus the following marginal of the
Wigner function 
\[
\sigma \left( \mathbf{p}\right) =\int W\left( \mathbf{x,p}\right) d^{3}x%
\mathbf{\geq }0, 
\]
may correspond to information about the expected future motion of the
physical particle. Indeed the distribution $\sigma \left( \mathbf{p}\right) $
agrees with the quantum ``probability distribution'' of momenta. These
arguments suggest why the Wigner function should not be understood as a
probability distribution in phase space. This fits in the fact that
(quantum) Heisenberg uncertainty relations forbid the existence of
simultaneous well defined position and momentum.

In summary we cannot assume that the quantum state of a particle, say an
electron, can be defined by position and momentum as is the case in
classical mechanics. As said above a physical electron is not a point (or
small) particle, but a complex system consisting of a cloud of electrons and
positrons interacting with electromagnetic interacting with the vacuum
radiation (ZPF) and possibly other fields. I will discuss the subject again
in section 8.1.

\subsection{The Weyl-Wigner formalism in quantum field theory}

The Weyl-Wigner transform has been trivially extended to the radiation field
provided we interpret $\hat{x}_{j}$ and $\hat{p}_{j}$ to be the sum and the
difference of the socalled creation, $\hat{a}_{j}^{\dagger },$ and
annihilation, $\hat{a}_{j},$ operators of the $j$ normal mode of the field 
\cite{Frontiers}, \cite{FOOP}. That is

\begin{eqnarray*}
\hat{x}_{j} &\equiv &\frac{c}{\sqrt{2}\omega _{j}}\left( \hat{a}_{j}+\hat{a}%
_{j}^{\dagger }\right) ,\hat{p}_{j}\equiv \frac{i 
\rlap{\protect\rule[1.1ex]{.325em}{.1ex}}h%
\omega _{j}}{\sqrt{2}c}\left( \hat{a}_{j}-\hat{a}_{j}^{\dagger }\right) \\
&\Rightarrow &\hat{a}_{j}=\frac{1}{\sqrt{2}}\left( \frac{\omega _{j}}{c}\hat{%
x}_{j}+\frac{ic}{
\rlap{\protect\rule[1.1ex]{.325em}{.1ex}}h%
\omega _{j}}\hat{p}_{j}\right) ,\hat{a}_{j}^{\dagger }=\frac{1}{\sqrt{2}}%
\left( \frac{\omega _{j}}{c}\hat{x}_{j}-\frac{ic}{%
\rlap{\protect\rule[1.1ex]{.325em}{.1ex}}h%
\omega _{j}}\hat{p}_{j}\right) .
\end{eqnarray*}
Here $
\rlap{\protect\rule[1.1ex]{.325em}{.1ex}}h%
$ is Planck constant, $c$ the velocity of light and $\omega _{j}$ the
frequency of the normal mode. In the following I will use units $
\rlap{\protect\rule[1.1ex]{.325em}{.1ex}}h%
=c=1$, but these parameters will be restored in some cases. For the sake of
clarity I shall represent the operators in HS with a `hat', e. g. $\hat{a}%
_{j},\hat{a}_{j}^{\dagger },$ and the amplitudes in the Wigner
representation without `hat', e. g. $a_{j},a_{j}^{*}.$

The transform provides an operator $\hat{f}$ , written in terms of operators 
$\hat{a}_{j}$ and $\hat{a}_{j\text{ }}^{\dagger }$, from any function $f$ of
the complex amplitudes $a_{j}$ and $a_{j}^{*}$ as follows 
\begin{eqnarray}
\hat{f} &=&\frac{1}{(2\pi ^{2})^{2n}}\prod_{j=1}^{n}\{\int_{-\infty
}^{\infty }d\lambda _{j}\int_{-\infty }^{\infty }d\mu _{j}\exp \left[
i\lambda _{j}\left( \hat{a}_{j}+\hat{a}_{j}^{\dagger }\right) +\mu
_{j}\left( \hat{a}_{j}-\hat{a}_{j}^{\dagger }\right) \right]  \nonumber \\
\; &&\int_{-\infty }^{\infty }d\mathrm{Re}a_{j}\int_{-\infty }^{\infty }d%
\mathrm{Im}a_{j}f\left( \left\{ a_{j},a_{j}^{*}\right\} \right) \exp
[-2i(\lambda _{j}\mathrm{Re}a_{j}+\mu _{j}\mathrm{Im}a_{j})]\}.  \label{h1}
\end{eqnarray}
The transform is \textit{invertible} that is 
\begin{eqnarray}
f\left( \left\{ a_{j},a_{j}^{*}\right\} \right) &=&\frac{1}{(2\pi ^{2})^{n}}%
\prod_{j=1}^{n}\int_{-\infty }^{\infty }d\lambda _{j}\int_{-\infty }^{\infty
}d\mu _{j}\exp \left[ -2i\lambda _{j}\mathrm{Re}a_{j}-2i\mu _{j}\mathrm{Im}%
a_{j}\right]  \nonumber \\
&&\times Tr\left\{ \hat{f}\exp \left[ i\lambda _{j}\left( \hat{a}_{j}+\hat{a}%
_{j}^{\dagger }\right) +\mu _{j}\left( \hat{a}_{j}-\hat{a}_{j}^{\dagger
}\right) \right] \right\} .  \label{h2}
\end{eqnarray}
The transform is\textit{\ linear}, that is if $f$ is the transform of $\hat{f%
}$ and $g$ the transform of $\hat{g}$, then the transform of $\hat{f}$ +$%
\hat{g}$ is $f+g$. Other properties may be seen in the references\cite
{Frontiers}, \cite{FOOP}.

Getting the field operators associated to given field amplitudes or to
obtain the amplitudes from the operators is straigtforward taking eqs.$%
\left( \ref{h1}\right) $ or $\left( \ref{h2}\right) $ into account.
Particular instances are the following 
\begin{eqnarray}
\hat{a}_{j} &\leftrightarrow &a_{j},\hat{a}_{j}^{\dagger }\leftrightarrow
a_{j}^{*},\frac{1}{2}\left( \hat{a}_{j}^{\dagger }\hat{a}_{j}+\hat{a}_{j}%
\hat{a}_{j}^{\dagger }\right) \leftrightarrow a_{j}a_{j}^{*}=\left|
a_{j}\right| ^{2},  \nonumber \\
\hat{a}_{j}^{\dagger }\hat{a}_{j} &=&\frac{1}{2}\left( \hat{a}_{j}^{\dagger }%
\hat{a}_{j}+\hat{a}_{j}\hat{a}_{j}^{\dagger }\right) +\frac{1}{2}\left( \hat{%
a}_{j}^{\dagger }\hat{a}_{j}-\hat{a}_{j}\hat{a}_{j}^{\dagger }\right)
\leftrightarrow \left| a_{j}\right| ^{2}-\frac{1}{2},  \nonumber \\
\hat{a}_{j}\hat{a}_{j}^{\dagger } &\leftrightarrow &\left| a_{j}\right| ^{2}+%
\frac{1}{2},\hat{a}_{j}{}^{2}\rightarrow a_{j}^{2},\hat{a}_{j}^{\dagger
2}\rightarrow \hat{a}_{j}^{*2},  \nonumber \\
\left( \hat{a}_{j}^{\dagger }+\hat{a}_{j}\right) ^{n} &\leftrightarrow
&\left( a_{j}+a_{j}^{*}\right) ^{n},\left( \hat{a}_{j}^{\dagger }-\hat{a}%
_{j}\right) ^{n}\leftrightarrow \left( a_{j}-a_{j}^{*}\right) ^{n},n\text{
an integer.}  \label{2}
\end{eqnarray}
I stress that the quantities $a_{j}$ and $a_{j}^{*}$ are c-numbers and
therefore they commute with each other. As said above it is standard in HS
to call $\left\{ \hat{a}_{j}\right\} $ and $\left\{ \hat{a}_{j}^{\dagger
}\right\} $ the annihilation and creation operators of photons,
respectively. We will use these names here for clarity although our study
will not introduce ``photons'' at any stage. The first two eqs.$\left( \ref
{2}\right) $ mean that the Weyl transform eq.$\left( \ref{h2}\right) $ in
expressions \textit{linear in creation and/or annihilation operators} just
implies ``\textit{removing the hats}''. However this is not the case in
nonlinear expressions in general. In fact from the latter two eqs.$\left( 
\ref{2}\right) $ plus the linearity property it follows that for a product
in the Weyl formalism the HS counterpart is 
\begin{equation}
a_{j}^{k}a_{j}^{*^{l}}\leftrightarrow (\hat{a}_{j}^{k}\hat{a}_{j}^{\dagger
l})_{sym},  \label{2b}
\end{equation}
where the subindex $sym$ means that the term is actually a sum of products
of the operators involved, written in all possible orderings, divided by the
number of terms. Hence the stochastic field amplitudes corresponding to a
product of field operators may be easily obtained transforming firstly the
product of operators into a sum of terms presenting a symmetrical order.
This may be always achieved taking the commutation rules eqs.$\left( \ref{h3}%
\right) $ into account.

In analogy with particle mechanics the Weyl-Wigner transform eq.$\left( \ref
{h3}\right) $ of the density operator (or density matrix) representing a
quantum state of radiation may be named Wigner function of the state. In
particular eq.$\left( \ref{f3}\right) $ is the Wigner function of the vacuum
state. Therefore I will name WW the formalism obtained from HS via the Weyl
transform eq.$\left( \ref{h2}\right) $, which is currently named Wigner
representation. It admits an interpretation in terms of random variables and
stochastic processes provided we respect some constraints. In particular the
radiation Wigner function may be interpreted as a probability distribution
only if it is non-negative.

The evolution of the free field given in eq.$\left( \ref{f5}\right) $ may be
interpreted saying that the amplitude of every normal mode fulfils 
\begin{equation}
a_{j}\left( t\right) =a_{j}\left( 0\right) \exp \left( \mathbf{-}i\omega
_{j}t\right) ,a_{j}^{*}\left( t\right) =a_{j}^{*}\left( 0\right) \exp \left(
i\omega _{j}t\right) .  \label{8h}
\end{equation}
The evolution of the field operators in HS, obtained from the Weyl
transform, is quite similar. It is interesting that it may be derived in
general via a commutator involving the HS Hamiltonian, as is well known. In
the particular case of eq.$\left( \ref{8h}\right) $ we have 
\begin{equation}
\frac{d\hat{a}_{k}}{dt}=-\frac{i}{
\rlap{\protect\rule[1.1ex]{.325em}{.1ex}}h%
}\left[ \hat{H},\hat{a}_{k}\right] ,\hat{H}=\frac{1}{2}\sum_{j} 
\rlap{\protect\rule[1.1ex]{.325em}{.1ex}}h%
\omega _{j}\left( \hat{a}_{j}\hat{a}_{j}^{\dagger }+\hat{a}_{j}^{\dagger }%
\hat{a}_{j}\right) .  \label{h9}
\end{equation}
This illustrates the connection, via Weyl-Wigner transform, between the
classical (Maxwell) evolution eq.$\left( \ref{f1}\right) $ and the HS
Heisenberg equation of motion eq.$\left( \ref{h9}\right) $ in a particular
case. For the general treatment of evolution in the WW formalism see the
references \cite{Frontiers}, \cite{FOOP}.

Eqs.$\left( \ref{h1}\right) $ or $\left( \ref{h2}\right) $ are suited in
order to transform observables, represented by density matrices in the HS
formalism, into functions of the field amplitudes in the WW formalism.
However the density operators not always are written as functions of the
operators $\left\{ \hat{a}_{j},\hat{a}_{j}^{\dagger }\right\} $, and the
Weyl transform requires a more sophisticated treatment as show in the
following.

\subsection{States of radiation in Weyl-Wigner and Hilbert space formalisms}

The Weyl-Wigner transform should be written for \textit{the total amplitudes}
of the modes, that for clarity we will label $\left\{
c_{j},c_{j}^{*}\right\} $ as in eq.$\left( \ref{f9}\right) $ in order to
avoid any confussion with $\left\{ a_{j},a_{j}^{*}\right\} $ used above for
the particular case of the vacuum state. Thus we shall rewrite eq.$\left( 
\ref{h2}\right) $ as follows (for a single mode) 
\begin{eqnarray}
f\left( c,c^{*}\right) &=&\frac{1}{2\pi ^{2}}\int_{-\infty }^{\infty
}d\lambda \int_{-\infty }^{\infty }d\mu \exp \left[ -2i\lambda \mathrm{Re}%
c-2i\mu \mathrm{Im}c\right]  \nonumber \\
&&\times Tr\left\{ \hat{f}\exp \left[ i\lambda \left( \hat{a}+\hat{a}%
^{\dagger }\right) +\mu \left( \hat{a}-\hat{a}^{\dagger }\right) \right]
\right\} .  \label{h11}
\end{eqnarray}

A relevant application of the WW transform is to get the operator, $\hat{v}$%
, associated to the vacuum state in HS. This operator will be the WW
transform of the stochastic vacuum distribution eq.$\left( \ref{f3}\right) $%
. I show from eq.$\left( \ref{h11}\right) $ that the solution is

\begin{equation}
\hat{v}=\left| 0><0\right| ,  \label{4h}
\end{equation}
where $\mid 0\rangle $ is named ``vacuum state vector'' that fulfils 
\begin{equation}
\hat{a}_{j}\mid 0\rangle =\langle 0\mid \hat{a}_{j}^{\dagger }=0,  \label{h4}
\end{equation}
$0$ meaning the nul vector in the Hilbert space. We shall do the proof for a
single mode, taking the Campbell-Haussdorf formula into account, that is 
\begin{equation}
\exp \left( \hat{A}+\hat{B}\right) =\exp \left( \hat{A}\right) \exp \left( 
\hat{B}\right) \exp \left( -\frac{1}{2}\left[ \hat{A},\hat{B}\right] \right)
,  \label{h5}
\end{equation}
valid if the operator $\left[ \hat{A},\hat{B}\right] $ commutes with $\hat{A}
$ and with $\hat{B}$. Hence the trace involved in eq.$\left( \ref{h2}\right) 
$ becomes 
\begin{eqnarray}
&&Tr\left\{ \left| 0><0\right| \exp \left[ i\lambda \left( \hat{a}+\hat{a}%
^{\dagger }\right) +\mu \left( \hat{a}-\hat{a}^{\dagger }\right) \right]
\right\}  \nonumber \\
&=&\langle 0\mid \exp \left[ (i\lambda -\mu )\hat{a}^{\dagger }\right] \exp
\left[ (i\lambda +\mu )\hat{a}\right] \mid 0\rangle \exp \left[ -\lambda
^{2}/2-\mu ^{2}/2\right]  \nonumber \\
&=&\exp \left[ -\lambda ^{2}/2-\mu ^{2}/2\right] .  \label{h6}
\end{eqnarray}
If this is inserted in eq.$\left( \ref{h2}\right) $ we get, for every mode, 
\begin{eqnarray}
f\left( a,a^{*}\right) &=&\frac{1}{2\pi ^{2}}\int_{-\infty }^{\infty
}d\lambda \int_{-\infty }^{\infty }d\mu \exp \left[ -2i\lambda \mathrm{Re}%
a-2i\mu \mathrm{Im}a-\frac{1}{2}\left( \lambda ^{2}+\mu ^{2}\right) \right] 
\nonumber \\
&=&\frac{1}{\pi }\exp \left[ -2\left( \mathrm{Re}a\right) ^{2}-2\left( 
\mathrm{Im}a\right) ^{2}\right] ,  \label{h7}
\end{eqnarray}
that agrees with eq.$\left( \ref{f3}\right) .$

Another example is the HS state corresponding to the WW state eq.$\left( \ref
{f9}\right) .$ In this case the distribution function $f\left(
c,c^{*}\right) $ is given by eq.$\left( \ref{f9}\right) $ and we must find
the associated density matrix $\hat{f}.$ In eq.$\left( \ref{4h}\right) $ I
propose that the solution is 
\begin{equation}
\hat{f}=\left| s><s\right| ,  \label{10h}
\end{equation}
where $\mid s\rangle $ is defined by 
\begin{equation}
\hat{c}\mid s\rangle =\left( \hat{a}+b\right) \mid s\rangle =0,\langle s\mid 
\hat{c}^{\dagger }=\langle s\mid \left( \hat{a}^{\dagger }+b^{*}\right) =0,
\label{h10}
\end{equation}
by analogy with eq.$\left( \ref{9f}\right) .$ I point out that $b$ is a
complex c-number (or more properly a number times the unit operator in the
Hilbert space) and $\hat{a}$ is the standard annihilation operator
fulfilling eq.$\left( \ref{h4}\right) .$ Then steps similar to those leading
to eq.$\left( \ref{h6}\right) $ give 
\begin{eqnarray}
&&Tr\left\{ \left| s><s\right| \exp \left[ i\lambda \left( \hat{a}+\hat{a}%
^{\dagger }\right) +\mu \left( \hat{a}-\hat{a}^{\dagger }\right) \right]
\right\}  \nonumber \\
&=&\langle s\mid \exp \left[ (i\lambda -\mu )\hat{a}^{\dagger }\right] \exp
\left[ (i\lambda +\mu )\hat{a}\right] \mid s\rangle \exp \left[ -\lambda
^{2}/2-\mu ^{2}/2\right]  \nonumber \\
&=&\exp \left[ -\lambda ^{2}/2-\mu ^{2}/2\right] \exp \left[ -(i\lambda -\mu
)b^{*}\right] \exp \left[ -(i\lambda +\mu )b\right] .  \label{11h}
\end{eqnarray}
Inserting this in eq.$\left( \ref{h7}\right) $ and performing the $\lambda $
and $\mu $ integrals we obtain eq.$\left( \ref{f9}\right) ,$ that confirms
that eq.$\left( \ref{10h}\right) $ is indeed the HS density matrix
representative of the state eq.$\left( \ref{f9}\right) .$ It is interesting
that the vector state eq.$\left( \ref{h10}\right) $ is named a coherent
state of the radiation, characterized by the amplitude -$b,$ which fulfils
the equation 
\[
\hat{a}\mid s\rangle =-b\mid s\rangle . 
\]

As pointed out above not all density matrices that people assume to
correspond to states in HS are actually physical states. In particular this
happens in all instances where the radiation Wigner function is not
positive, e.g. all states with $n$ photons, $n\neq 0.$ On the other hand all
(positive) probability distributions of the amplitudes $\left\{
a_{j},a_{j}^{*}\right\} $ might be considered possible states of the
(stochastic) radiation field, but it may be that many of them do not exist
in nature and cannot be manufactured in the laboratory. In any case I stress
that \textit{the realistic interpretation of quantum theory that we are
searching for does not require getting an intuitive picture of all states
and observables assumed in HS (}which is indeed impossible\textit{) but to
understand actual or feasible experiments.}

\subsection{Expectation values}

Expectation values may be calculated in the WW formalism as follows. In the
HS formalism they read $Tr(\hat{\rho}\hat{M})$, or in particular $\langle
\psi \mid \hat{M}\mid \psi \rangle ,$ whence the translation to the WW
formalism is obtained taking into account that the trace of the product of
two operators becomes 
\begin{mathletters}
\begin{equation}
Tr(\hat{\rho}\hat{M})=\int W_{\hat{\rho}}\left\{ \hat{a}_{j},\hat{a}%
_{j}^{\dagger }\right\} W_{\hat{M}}\left\{ \hat{a}_{j},\hat{a}_{j}^{\dagger
}\right\} \prod_{j}d\mathrm{Re}a_{j}d\mathrm{Im}a_{j}.  \label{trace}
\end{equation}
That integral is the WW counterpart of the trace operation in the HS
formalism.

Particular instances are the following expectations that will be of interest
later 
\end{mathletters}
\begin{eqnarray}
\left\langle \left| a_{j}\right| ^{2n}\right\rangle &\equiv &\int d\Gamma
W_{0}\left| a_{j}\right| ^{2n}=\frac{n!}{2^{n}},\left\langle
a_{j}^{n}a_{k}^{*m}\right\rangle =\delta _{jk}\delta _{mn}\left\langle
\left| a_{j}\right| ^{2n}\right\rangle =\frac{n!}{2^{n}},  \nonumber \\
\left\langle \left| a_{j}\right| ^{2n}\left| a_{k}\right| ^{2m}\right\rangle
&=&\left\langle \left| a_{j}\right| ^{2n}\right\rangle \left\langle \left|
a_{k}\right| ^{2m}\right\rangle \text{ if }j\neq k.  \nonumber \\
\left\langle 0\left| \hat{a}_{j}^{\dagger }\hat{a}_{j}\right| 0\right\rangle
&=&\int d\Gamma (a_{j}^{*}a_{j}-\frac{1}{2})W_{0}=0,  \nonumber \\
\left\langle 0\left| \hat{a}_{j}\hat{a}_{j}^{\dagger }\right| 0\right\rangle
&=&\int d\Gamma (a_{j}^{*}a_{j}+\frac{1}{2})W_{0}=2\left\langle \left|
a_{j}\right| ^{2}\right\rangle =1,  \label{a3}
\end{eqnarray}
where $W_{0}$ is the Wigner function of the vacuum, eq.$\left( \ref{f3}%
\right) $. This means that in the WW formalism the field amplitude $a_{j}$
(coming from the vacuum) behaves like a complex random variable with
Gaussian distribution and mean square modulus $\left\langle \left|
a_{j}\right| ^{2}\right\rangle =1/2.$ I point out that the integral for any
mode not entering in the function $M\left( \left\{ a_{j},a_{j}^{*}\right\}
\right) $ of eq.$\left( \ref{trace}\right) $ gives unity in the integration
due to the normalization of the Wigner function eq.$\left( \ref{f3}\right) $%
. An important consequence of eq.$\left( \ref{a3}\right) $ is that \textit{%
normal (antinormal) ordering of one creation and one annihilation operators
in the Hilbert space formalism becomes subtraction (addition) of 1/2 to the
field intensity in the WW formalism. The normal ordering rule is equivalent
to subtracting the vacuum contribution }as said above.

\section{Entanglement and Bell inequalities}

In this section I comment on two related difficulties that allegedly prevent
a realistic interpretation of quantum theory, namely the non-classical
properties of entangled states and the empirical violation of Bell
inequalities. I shall show that the first difficulty may be removed if we
assume that the quantum vacuum fields are real stochastic fields. In
particular I shall study the relevance of the vacuum electromagnetic field
in quantum optics. For the empirical violation of Bell inequalities I have
no clear interpretation, and it will not be discussed in this article.

\subsection{Entanglement}

Entanglement is a quantum property that may be easily defined within the HS
formalism, but the definition does not provide any intuitive picture. It
appears in systems with several degrees of freedom when the total state
vector of the system cannot be written as a product of vectors associated to
one degree of freedom each. In formal terms a typical entangled state is the
following 
\begin{equation}
\mid \psi \left( 1,2\right) \rangle =\sum_{m,n}c_{mn}\mid \psi _{m}\left(
1\right) \rangle \mid \psi _{n}\left( 2\right) \rangle ,  \label{entangled}
\end{equation}
where $1$ and $2$ correspond to two different degrees of freedom, usually
belonging to different subsystems that may be placed far from each other,
and $c_{mn}$ are complex numbers. The essential condition is that the state
eq.$\left( \ref{entangled}\right) $ cannot be written as a single product,
that is the sum cannot be reduced to just one term via a change of basis in
the Hilbert space. Entanglement appears as a specifically quantum form of
correlation, which is claimed to be dramatically different from the
correlations that appear in all other branches of science, including
classical physics.

The relevance of entanglement was stressed in 1935 by Schr\"{o}dinger \cite
{Schrodinger} 1935, who wrote that ``it is not one but \textit{the }%
characteristic trait of quantum mechanics''. He also pointed out the
difficulty to understand entanglement with his celebrated example of the cat
suspended between life and death. Indeed if one assumes that quantum
mechanics is complete, i.e. that a state-vector like eq.$\left( \ref
{entangled}\right) $ represents a pure state, then a realistic
interpretation is impossible because we are confronted with consequences in
sharp contradiction with both the intuition and a well established pardigm,
namely that complete information about the whole requires complete
information about every part. In fact we are compelled to believe that a
state-vector like eq.$\left( \ref{entangled}\right) $ represents complete
information about the state of the system but incomplete information about
every one of the subsystems. Indeed according to quantum theory the state of
the first subsystem should be obtained by taking the partial trace with
respect to the second subsystem, leading to the following density matrix
(assuming all state-vectors normalized) 
\begin{equation}
\rho \left( 1\right) =Tr_{2}\left[ \mid \psi \left( 1,2\right) \rangle
\langle \psi \left( 1,2\right) \mid \right] =\sum_{m}\left| c_{m}\right|
^{2}\mid \psi _{m}\left( 1\right) \rangle \langle \psi _{m}\left( 1\right)
\mid .  \label{mixed}
\end{equation}
The density matrix represents a mixed state, where the information is
incomplete, that is we only know the probabilities, $P_{m}=\left|
c_{m}\right| ^{2},$ for the first subsystem to be in the different states $%
\mid \psi _{m}\left( 1\right) \rangle $.

An important result is that entanglement is a necessary condition for the
violation of a Bell inequalities \cite{Santosajp}.

\subsection{Bell inequalities}

It is common wisdom that any correlation between two events, say $A$ and $B$%
, is either a causal connection or it derives from a common cause. There is
causal connection if $A$ is the cause of $B$ or $B$ the cause of $A$, and a
common cause means that there is another event $C$ that causes both $A$ and $%
B$. In formal terms we may write either $A\Rightarrow B$ or $B\Rightarrow A$
for causal connection, $C\Rightarrow A$ and\ $C\Rightarrow B$ for common
cause. In 1965 John Bell allegedly proved that the said common wisdom is not
true according to quantum mechanics. In fact he derived inequalities\cite
{Bell} that he claimed to be necessary conditions for the existence of a
common cause, and pointed out possible experiments where the inequalities
would be violated.

Typical experimental tests of the Bell inequalities consist of preparing a
system that produces pairs of signals, one of them going to an observer
Alice and the other one to observer Bob. Alice may measure a dichotomic
property $a_{1}$ on her signal with the possible results $\left\{
0,1\right\} ,$ and in another run of the experiment she may measure $a_{2}$
also with the possible results $\left\{ 0,1\right\} .$ Similarly Bob may
measure either $b_{1}$ or $b_{2}$ with the possible results $\left\{
0,1\right\} .$ Alice and Bob may perform coincidence measurements of $a_{j}$
and $b_{k}.$ After many runs of the experiment with identical preparations
of the system, Alice may obtain from the frequencies the single probability, 
$P\left( a_{j}\right) $, that the result in her measurement is $1$, and
similarly Bob may got the probability $P\left( b_{k}\right) .$ They may also
obtain the probability $P\left( a_{j}b_{k}\right) $ that both results are $1$
in a coincidence measurement. Then the following Bell inequality\cite{CH} 
\begin{equation}
P\left( a_{2}\right) +P\left( b_{1}\right) \geq P\left( a_{1}b_{1}\right)
+P\left( a_{2}b_{1}\right) +P\left( a_{2}b_{2}\right) -P\left(
a_{1}b_{2}\right)  \label{CH}
\end{equation}
should hold true.

The relevant fact is that quantum mechanics predicts violations of Bell
inequalities in some cases. The contradiction has been named ``Bell\'{}s
theorem'': \textit{quantum mechanics is not compatible with local realism}.
Local realism is the assertion that all correlations in nature are either
causal connections or derive from a common cause. The word ``local'' is
introduced because a direct communication between Alice and Bob could
produce results violating eq.$\left( \ref{CH}\right) $ which would
invalidate the test. The possible communication is named local if any
possible information travel with velocity not greater than the speed of
light, whence locality should be better named ``relativistic causality''. As
a consequence the crucial experiments must be performed so that coincidence
measurements by Alice and Bob take place both within a time window $\Delta t$
smaller than the distance between their measuring devices divided by the
velocity of light, that is with spacelike separation in the sense of
relativity theory.

Many experiments have been performed in the last 50 years in order to test
Bell inequalities with results that generally agree with quantum
predictions, but there are loopholes for the proof that local realism is
refuted. In particular in most of the performed experiments the spacelike
separation is not guaranteed. The reader should consult the vast literature
on the subject. See e.g. \cite{Santosajp},\cite{RMP}.

In the last decades most tests of the inequalities have used entangled
photon pairs produced via spontaneous parametric down conversion (SPDC). In
section 7.6 I shall analyze a representative test similar to those providing
for the first time the loophole-free violation of a Bell inequality\cite
{Shalm},\cite{Giustina}. The empirical violation is interpreted as a
refutation of local realism.

\subsection{Spontaneous parametric down conversion (SPDC)}

SPDC has been the main source of entangled photon pairs from about 1980. In
the following I will study, within the quantum Hilbert space formalism (HS),
the SPDC process and a simple experiment involving entangled photon pairs. I
shall work in the Heisenberg picture where the obvervables evolve, see eq.$%
\left( \ref{h19}\right) $ below, but the state vector is fixed, in our case
the vacuum state $\mid 0\rangle .$ In section 7.4 I shall pass to the WW
formalism, which suggests an interpretation of SPDC experiments in terms of
random variables and stochastic processes without any reference to photons.

SPDC is produced when a pumping laser impinges a crystal possessing
nonlinear electric susceptibility. Radiation with several colors may be
observed going out from the opposite side of the crystal. By means of
appropriate apertures two beams of the radiation may be selected, which in
quantum language consist of a set of entangled photon pairs, one photon of
every pair in each beam.

The HS theory of the process is as follows, with the simplification of
taking only two radiation modes into account, having amplitudes $\hat{a}_{s},%
\hat{a}_{i}$. Avoiding a detailed study of the physics inside the crystal,
that may be seen elsewhere \cite{Dechoum}, \cite{MS}, we might describe the
phenomenon with a model interaction Hamiltonian \cite{Ou}, that is 
\begin{equation}
\hat{H}_{I}=A\hat{a}_{s}^{\dagger }\hat{a}_{i}^{\dagger }\exp \left(
-i\omega _{P}t\right) +A^{*}\hat{a}_{s}\hat{a}_{i}\exp \left( i\omega
_{P}t\right) ,  \label{50}
\end{equation}
when the laser is treated as a classically prescribed, undepleted and
spatially uniform field of frequency $\omega _{P}.$ The interaction of the
pumping laser with the incoming vacuum mode, $\hat{a}_{s},$ within the
crystal produces a new field with amplitude $D\hat{a}_{s}^{\dagger },$ named
``signal''. If the beams have been adequately chosen, that signal travels
superposed to the vacuum field $\hat{a}_{i}$ after exiting the crystal.
Similarly the vacuum field $\hat{a}_{i}$ produces a field $D\hat{a}%
_{i}^{\dagger },$ named ``idler'', that travels superposed to the vacuum
field $\hat{a}_{s}$.

As a result the radiation fields at the crystal exit may be represented by 
\begin{eqnarray}
\hat{a}_{s}(\mathbf{r,}t) &=&\left[ \hat{a}_{s}(0)+D\hat{a}_{i}^{\dagger
}\left( 0\right) \right] \exp \left( i\mathbf{k}_{s}\mathbf{\cdot r}-i\omega
_{s}t\right) ,  \nonumber \\
\hat{a}_{i}(\mathbf{r,}t) &=&\left[ \hat{a}_{i}(0)+D\hat{a}_{s}^{\dagger
}\left( 0\right) \right] \exp \left( i\mathbf{k}_{i}\mathbf{\cdot r}-i\omega
_{i}t\right) ,  \label{h19}
\end{eqnarray}
where the wavevectors $\mathbf{k}_{s}$ and $\mathbf{k}_{i}$ form a finite
angle amongst them.The parameter $D$ is proportional to the interaction
coefficient $A$ eq.$\left( \ref{50}\right) $ and it depends also on the
crystal size. In practice it fulfils $\left| D\right| <<1.$ The following
equality holds for the frequencies of the selected beams 
\begin{equation}
\omega _{P}=\omega _{s}+\omega _{i},  \label{h20}
\end{equation}
which is usually interpreted assuming that the signal and idler photons,
with energies $
\rlap{\protect\rule[1.1ex]{.325em}{.1ex}}h%
\omega _{s}$ and $
\rlap{\protect\rule[1.1ex]{.325em}{.1ex}}h%
\omega _{i},$ were the result of the division of one laser photon with
energy $
\rlap{\protect\rule[1.1ex]{.325em}{.1ex}}h%
\omega _{P}$. That is eq.$\left( \ref{h20}\right) $ is viewed as ``energy
conservation'' in the splitting of laser photons. However I interpret it as
a condition of frequency matching, induced by the nonlinear susceptibility,
with no reference to photons.

In the following I will ignore the spacetime dependence, whence eqs.$\left( 
\ref{h19}\right) $ will be written 
\begin{equation}
\hat{E}_{A}^{+}=\hat{a}_{s}+D\hat{a}_{i}^{\dagger },\hat{E}_{B}^{+}=\hat{a}%
_{i}+D\hat{a}_{s}^{\dagger }.  \label{h21}
\end{equation}
These equations are the formal representation of entangled photon pairs in
the Heisenberg picture of the HS formalism, and they show a strong
correlation between the fields $\hat{E}_{A}^{+}$ and $\hat{E}_{B}^{+}$.

As a simple application I shall derive the quantum prediction for an
experiment that consists of measuring the single and coincidence detection
rates when the beams with fields eqs.$\left( \ref{h21}\right) $ arrive at
Alice and Bob detectors, respectively. It is convenient to get the quantum
prediction in terms of the probability of detection, $P$, in a given time
window. Thus if we divide the unit of time in a number $n$ of windows the
detection rate would be $R=nP$. In the quantum HS formalism Alice single
detection probability is given by the following vacuum expectation (to order 
$O\left( \left| D\right| ^{2}\right) )$ 
\begin{eqnarray}
P_{A} &=&\left\langle 0\left| \hat{E}_{A}^{-}\hat{E}_{A}^{+}\right|
0\right\rangle =\left\langle 0\left| \left[ \hat{a}_{s}^{\dagger }+D^{*}\hat{%
a}_{i}\right] \left[ \hat{a}_{s}+D\hat{a}_{i}^{\dagger }\right] \right|
0\right\rangle  \nonumber \\
&=&\left\langle 0\left| \hat{a}_{s}^{\dagger }\hat{a}_{s}\right|
0\right\rangle +\left| D\right| ^{2}\left\langle 0\left| \hat{a}_{i}\hat{a}%
_{i}^{\dagger }\right| 0\right\rangle =\left| D\right| ^{2},  \label{h22}
\end{eqnarray}
where only one out of four terms contributes, but I have written explicitly
two of them for clarity. And similar for Bob. It is easy to prove that the
spacetime factors, explicit in eq.$\left( \ref{h19}\right) ,$ cancel.

The quantum prediction for the coincidence detection probability is 
\[
P_{AB}=\frac{1}{2}\left\langle 0\left| \hat{E}_{A}^{-}\hat{E}_{B}^{-}\hat{E}%
_{B}^{+}\hat{E}_{A}^{+}\right| 0\right\rangle +\frac{1}{2}\left\langle
0\left| \hat{E}_{B}^{-}\hat{E}_{A}^{-}\hat{E}_{A}^{+}\hat{E}_{B}^{+}\right|
0\right\rangle . 
\]
In our case, taking into account that $\hat{E}_{i}^{+}$ and $\hat{E}_{s}^{+}$
commute, both terms are equal and we have 
\begin{eqnarray}
P_{AB} &=&\left\langle 0\left| \hat{E}_{A}^{-}\hat{E}_{B}^{-}\hat{E}_{B}^{+}%
\hat{E}_{A}^{+}\right| 0\right\rangle  \nonumber \\
&=&\left\langle 0\left| \left[ \hat{a}_{s}^{\dagger }+D^{*}\hat{a}%
_{i}\right] \left[ \hat{a}_{i}^{\dagger }+D^{*}\hat{a}_{s}\right] \left[ 
\hat{a}_{i}+D\hat{a}_{s}^{\dagger }\right] \left[ \hat{a}_{s}+D\hat{a}%
_{i}^{\dagger }\right] \right| 0\right\rangle  \nonumber \\
&=&\left| D\right| ^{2}\left\langle 0\left| \hat{a}_{i}\hat{a}_{i}^{\dagger }%
\hat{a}_{i}\hat{a}_{i}^{\dagger }\right| 0\right\rangle +O\left( \left|
D\right| ^{4}\right) =\left| D\right| ^{2}+O\left( \left| D\right|
^{4}\right) .\smallskip  \label{h23}
\end{eqnarray}
The quantum predictions eqs.$\left( \ref{h22}\right) $ and $\left( \ref{h23}%
\right) $ show that the correlation is the maximum possible, that is the
coincidence detection rate equals the single rate of either Alice or Bob. In
contrast if there was no correlation we should have $P_{AB}=$ $%
P_{A}P_{B}=\left| D\right| ^{4}<<\left| D\right| ^{2}$. In any case single
and detection probabilities obviously must fulfil $P_{AB}\leq P_{A}$ and $%
P_{AB}\leq P_{B}.$ In actual experiments the predictions for real detectors
should take into account the detection efficiency. If it is $\eta <1$ equal
for both detectors the prediction would be $P_{A}=P_{B}=\eta \left| D\right|
^{2},P_{AB}=\eta ^{2}\left| D\right| ^{2},$ confirmed in actual experiments.

Entanglement of the form eq.$\left( \ref{entangled}\right) $ may be
exhibited if we pass to the Schr\"{o}dinger picture, where the evolution
goes in the state. The appropriate representation of the joint quantum state
of the radiation at Alice and Bob detectors is 
\begin{equation}
\mid \psi \rangle =\sqrt{1-\left| D\right| ^{2}}\mid 0\rangle \mid 0\rangle
+\left| D\right| \mid 1\rangle \mid 1\rangle ,  \label{h24}
\end{equation}
which may be interpreted saying that the state of the radiation is entangled
and consists of two terms Alice and Bob having one photon each in the second
term and none of them having photons in the first term. I stress that in the
HS of quantum theory eq.$\left( \ref{h24}\right) $ represents a pure state,
not a statistical mixture. It cannot be interpreted as a probability $\left|
D\right| ^{2}$ of having two photons and a probability $1-\left| D\right|
^{2}$ of no photons. If $\hat{N}_{A}$ and $\hat{N}_{B}$ are the photon
number (operator) observables for Alice and Bob in a given time window, the
detection single probability will be 
\[
P_{A}=\langle \psi \mid \hat{N}_{A}\mid \psi \rangle =\langle 1\mid \left|
D\right| ^{2}\mid 1\rangle \langle 1\mid 1\rangle =\left| D\right| ^{2}, 
\]
and a similar for $P_{B}$. From the 2 terms of $\mid \psi \rangle $ eq.$%
\left( \ref{h24}\right) $ we get 4 terms for the expectation but 3 of them
do not contribute. The coincidence probability also consists of 4 terms, but
only one contributes, namely 
\[
P_{AB}=\langle \psi \mid \hat{N}_{A}\hat{N}_{B}\mid \psi \rangle =\langle
1\mid \hat{N}_{A}\mid 1\rangle \langle 1\mid \hat{N}_{B}\mid 1\rangle
=\left| D\right| ^{2}. 
\]
In summary eq.$\left( \ref{h24}\right) $ exhibits entanglement between the
vacuum and the two- photon state, as has been pointed out\cite{Milonni}.

\subsection{Stochastic interpretation of the correlation experiment}

The quantum-mechanical prediction for the experiment commented in the
previous section may be easily worked in the WW formalism. The Weyl
transform of the field operators eqs.$\left( \ref{h21}\right) $ are

\begin{equation}
E_{A}=a_{s}+Da_{i}^{*},E_{B}=a_{i}+Da_{s}^{*}.  \label{h27}
\end{equation}
Vacuum expectation in HS correspond in WW to an average weighted by the
vacuum probability distribution eq.$\left( \ref{f3}\right) .$ However the
detection probabilities in WW cannot be obtained just taking averages of eqs.%
$\left( \ref{h27}\right) ,$ but should be got from the Weyl transform of the
HS vacuum expectations. For Alice single detection probability the Weyl
transform of eq.$\left( \ref{h22}\right) $ is

\begin{equation}
P_{A}=\int d\Gamma (a_{s}^{*}a_{s}-\frac{1}{2})W_{0}+\left| D\right|
^{2}\int d\Gamma (a_{i}^{*}a_{i}+\frac{1}{2})W_{0}=\left| D\right| ^{2},
\label{h28}
\end{equation}
where eqs.$\left( \ref{a3}\right) $ have been taken into account. I ignore
two terms that do not contribute and are not relevant for the
interpretation. Similar for Bob detection probability $P_{B}$. The result
agrees with the prediction using HS, as it should because the WW formalism
is an equivalent form of quantum theory for the radiation field.

The different signs in front of 1/2 in the two terms of eq.$\left( \ref{h28}%
\right) $ may seem strange. Of course they appear in the Weyl transform of
eq.$\left( \ref{h22}\right) $ because the former comes from the vacuum
expectation of $\hat{a}_{s}^{\dagger }\hat{a}_{s}$ which is zero but the
latter from the vacuum expectation of $\hat{a}_{i}\hat{a}_{i}^{\dagger }$
which is unity. However in the WW formalism we are working with commuting
amplitudes and the different ordering should not make any difference. We may
understand intuitively the reason for the signs taking into account that the
second term of eq.$\left( \ref{h28}\right) $ corresponds to the signal (it
contains $\left| D\right| ^{2}$) but the first term corresponds to vacuum
modes that should not contribute to the detection and therefore should be
removed. The addition of 1/2 in the signal term effectively multiplies the
detection probability times 2. This is more difficult to understand
intuitively and I will not comment further on.

In order to derive the coincidence detection probability, $P_{AB}$, we might
proceed translating to the WW formalism the calculation made in section 7.3
using the HS formalism, which led to eq.$\left( \ref{h23}\right) $ (see \cite
{FOOP}). However I will not do that but make directly a stochastic
derivation of single and coincidence probabilities, which may allow
understanding more easily the physics of the experiment. I will start from
the fields eqs.$\left( \ref{h27}\right) $ and proceed using classical laws
and plausible assumptions for the correlations.

I shall start proposing a model of detection. According to our assumptions
any photodetector in free space is immersed in an extremely strong
stochastic radiation, infinite if no cut-off existed, see eq.$\left( \ref
{f3}\right) .$ Thus how might we explain that detectors are not activated by
the vacuum radiation? Firstly the strong vacuum field is effectively reduced
to a weaker level if \textit{we assume that only radiation within some
(small) frequency interval is able to activate a photodetector}, that is the
interval of sensitivity $\left( \omega _{1},\omega _{2}\right) $. Actually
the frequency selection is quite common in radiation detection, for instance
when tuning radio or TV. The theoretical explanation of this fact is easy,
that is detection takes place via resonance with some oscillator having the
same characteristic frequency than the radiation to be detected. For
instance an appropriate electric circuit in case of radiowaves or a
molecular resonator for visible light (e.g. molecules with a appropriate
frequency of excitation inside the elements of color vision in our retina).

However the problem is not yet solved because the signals involved in
experiments may have intensities of order the vacuum radiation in the said
frequency interval, whence the detector would be unable to distinguish a
signal from ZPF noise. Our assumption is that a \textit{detector may be
activated only when the Poynting vector (i.e. the directional energy flux)
of the incoming radiation is different from zero, including both signal and
vacuum fields}. To make a trivial comparison, we live immersed in air but
its pressure is almost unnoticed except when there is strong wind producing
an unbalanced force that pushes us towards a given direction.

Thus a plausible hypothesis is that \textit{light\ detectors possess an
active area, the probability of a photocount depending on the integrated
energy flux crossing that area during some activation time, T. }The
assumption allows understanding why the signals, but not the vacuum fields,
activate detectors. Indeed the ZPF arriving at any point (in particular the
detector) would be isotropic on the average, whence the associated energy
flux integrated over a large enough time would be very small because
fluctuations are averaged out. Therefore only the signal, which is
directional, would produce a large integrated energy flux during the
activation time, thus givin rise to photocounts. A problem remains because
the integrated flux would not be strictly zero. Indeed the integrated flux
during a time integral T, divided by T would go to zero when T$\rightarrow
\infty $. Hence we may predict the existence of some dark rate induced by
vacuum fluctuations even at zero Kelvin. In summary we are assuming that
photocounts are not produced by an instantaneous interaction of the
radiation field with the detector but the activation requires some time
interval, a fact well known by experimentalist.

After that I will obtain the detection probabilities as averages of
intensities derived from the fields eqs.$\left( \ref{h27}\right) $. I will
assume that the detection probability is proportional to the mean intensity
arriving at the detector, taking the proportionality coefficient as unit for
simplicity. For the single detection by Alice we get the detection
probability as the average of the intensity arriving at her detector, that
is 
\begin{equation}
P_{A}=\left\langle I_{A}+I_{A}^{ZPF}\right\rangle
,I_{A}=E_{A}E_{A}^{*}=\left| E_{s}\right| ^{2}.  \label{h29}
\end{equation}
According our previous analysis we should use time averages but we may
assume that they are equal to ensemble averages, a kind of ergodic property.
Eq.$\left( \ref{h29}\right) $ has two intensity contributions, the former $%
I_{A}$ coming from the signal and the latter $I_{A}^{ZPF}$ from the ZPF. On
the other hand if the laser pumping on the crystal was switch off, then the
total intensity arriving at Alice detector should be zero on the average,
that is 
\begin{equation}
\left\langle I_{A0}+I_{A}^{ZPF}\right\rangle =0\Rightarrow \left\langle
I_{A}^{ZPF}\right\rangle =-\left\langle I_{A0}\right\rangle ,  \label{h30}
\end{equation}
where $I_{A0}$ is the intensity arriving at the detector from the source in
place of the signal when there is no pumping$.$ The intensity $I_{A0}$ comes
from the vacuum fields and it may be derived from eqs.$\left( \ref{h27}%
\right) $ putting $D=0$. From eq.$\left( \ref{h30}\right) $ the probability
eq.$\left( \ref{h29}\right) $ becomes 
\[
P_{A}=\left\langle I_{A}\right\rangle -\left\langle I_{A0}\right\rangle , 
\]
where the second term correspond to the ZPF subtraction. This means that we
should not expect any detection if there is no signal, a quite plausible
result.

The radiation intensities may be obtained form the fields taking eqs.$\left( 
\ref{h27}\right) $ into account$,$ as follows 
\begin{equation}
I_{A}=\left| E_{A}\right| ^{2}=\left| a_{s}+Da_{i}^{*}\right|
^{2},I_{A0}=\left| E_{A0}\right| ^{2}=\left| a_{s}\right| ^{2}.  \label{h31}
\end{equation}
Eqs.$\left( \ref{h29}\right) $ to $\left( \ref{h31}\right) $ lead to 
\begin{equation}
P_{A}=\left\langle I_{A}-I_{A0}\right\rangle =\left\langle \left|
a_{s}+Da_{i}^{*}\right| ^{2}-\left| a_{s}\right| ^{2}\right\rangle
=\left\langle \left| Da_{i}^{*}\right| ^{2}\right\rangle =\frac{1}{2}\left|
D\right| ^{2},  \label{h32}
\end{equation}
where we take into account that 
\[
\left\langle Da_{s}a_{i}^{*}\right\rangle =\left\langle
D^{*}a_{s}^{*}a_{i}\right\rangle =0, 
\]
and we have calculated the expectation of $\left| a_{i}\right| ^{2}$ taking
the vacuum probability distribution eq.$\left( \ref{f3}\right) $ into
account. The same is obtained for Bob single detection.

The result eq.$\left( \ref{h32}\right) $ agrees with both the HS and WW
results, eqs.$\left( \ref{h22}\right) $ and $\left( \ref{h28}\right) ,$
except for a factor 1/2. It is caused by our choice, unity, for the
proportionality constant between field intensity and detection probability
made in eq.$\left( \ref{h29}\right) $.

The coincidence detection probability for a given time window will be the
average of the product of the field intensities whence the detection
probability per time window is obtained as follows 
\begin{equation}
P_{AB}=\left\langle \left( I_{A}+I_{A}^{ZPF}\right)
(I_{B}+I_{B}^{ZPF})\right\rangle .  \label{s7}
\end{equation}
As in the derivation of eq.$\left( \ref{h32}\right) ,$ the detection
probability $P_{AB}$ should be zero when the pumping is off, whence we get 
\begin{equation}
\left\langle \left( I_{A0}+I_{A}^{ZPF}\right)
(I_{B0}+I_{B}^{ZPF})\right\rangle =0.  \label{s70}
\end{equation}
From eqs.$\left( \ref{h30}\right) ,\left( \ref{s7}\right) $ and $\left( \ref
{s70}\right) $ and the plausible assumption that $I_{A}^{ZPF}$ and $%
I_{B}^{ZPF}$are uncorrelated with the signals we obtain 
\begin{equation}
P_{AB}=\left\langle I_{A}I_{B}\right\rangle -\left\langle
I_{A0}I_{B0}\right\rangle -\left\langle I_{A}\right\rangle \left\langle
I_{B0}\right\rangle -\left\langle I_{A0}\right\rangle \left\langle
I_{B}\right\rangle +2\left\langle I_{A0}\right\rangle \left\langle
I_{B0}\right\rangle .  \label{s71}
\end{equation}
The average of a single intensities in eq.$\left( \ref{h31}\right) $ may be
easily obtained taking the vacuum distribution eq.$\left( \ref{f3}\right) $
into account. We get 
\begin{equation}
\left\langle I_{A}\right\rangle =\left\langle I_{B}\right\rangle =\frac{1}{2}%
\left( 1+\left| D\right| ^{2}\right) ,\left\langle I_{A0}\right\rangle
=\left\langle I_{B0}\right\rangle =\frac{1}{2}.  \label{h33}
\end{equation}
For the average of products of intensities we have 
\begin{equation}
\left\langle I_{A0}I_{B0}\right\rangle =\left\langle \left| a_{s}\right|
^{2}\left| a_{i}\right| ^{2}\right\rangle =\left\langle \left| a_{s}\right|
^{2}\right\rangle \left\langle \left| a_{i}\right| ^{2}\right\rangle =\frac{1%
}{4},  \label{h34}
\end{equation}
\begin{eqnarray}
\left\langle I_{A}I_{B}\right\rangle &=&\left\langle \left|
a_{s}+Da_{i}^{*}\right| ^{2}\left| a_{i}+Da_{s}^{*}\right| ^{2}\right\rangle
\nonumber \\
&=&\langle [\left| a_{s}\right| ^{2}+\left| D\right| ^{2}\left| a_{i}\right|
^{2}+2\func{Re}\left( Da_{s}a_{i}^{*}\right) ]  \nonumber \\
&&[\left| a_{i}\right| ^{2}+\left| D\right| ^{2}\left| a_{s}\right| ^{2}+2%
\func{Re}\left( Da_{s}^{*}a_{i}\right) ]\rangle .  \label{3}
\end{eqnarray}
The terms with an odd number of amplitudes do not contribute (see eqs.$%
\left( \ref{a3}\right) )$ whence we get the following sum of averages to
order $\left| D\right| ^{2}$%
\begin{eqnarray}
\left\langle I_{A}I_{B}\right\rangle &=&\left\langle \left| a_{s}\right|
^{2}\left| a_{i}\right| ^{2}\right\rangle +\left| D\right| ^{2}\left\langle
\left| a_{s}\right| ^{4}+\left| a_{i}\right| ^{4}\right\rangle
+4\left\langle \func{Re}\left( Da_{s}a_{i}^{*}\right) \func{Re}\left(
Da_{s}^{*}a_{i}\right) \right\rangle  \nonumber \\
&=&\frac{1}{4}+\left| D\right| ^{2}+4\left\langle \func{Re}\left(
Da_{s}a_{i}^{*}\right) \func{Re}\left( Da_{s}^{*}a_{i}\right) \right\rangle .
\label{h}
\end{eqnarray}
We will show that the last term does not contribute whence, collecting all
terms, eq.$\left( \ref{s71}\right) $ becomes 
\begin{equation}
P_{AB}=\frac{1}{2}\left| D\right| ^{2}.  \label{h35}
\end{equation}

The reason why the last term of eq.$\left( \ref{h}\right) $ does not
contribute is that we cannot ignore the spacetime phase factors in this
case, see eq.$\left( \ref{h19}\right) .$ In fact $\func{Re}\left(
Da_{s}a_{i}^{*}\right) $ comes from the intensity arriving at Alice, but $%
\func{Re}\left( Da_{s}^{*}a_{i}\right) $ from the Bob intensity. In the
former we should include a phase $\exp \left( i\phi \right) $ and in the
latter $\exp \left( i\chi \right) $, these phases being uncorrelated.
Therefore in the average of the last term of eq.$\left( \ref{h}\right) $ the
phases give a nil contribution. In contrast all other terms contain absolute
values whence the phases disappear.

The results derived from a eqs.$\left( \ref{h32}\right) $ and $\left( \ref
{h35}\right) ,$ that have been obtained from the fields via our stochastic
approach, reproduce the relevant result of the experiment, namely that there
is a maximum positive correlation shown by the equality $P_{AB}=P_{A}=P_{B},$
which is also predicted by the HS results eqs.$\left( \ref{h22}\right) $ and 
$\left( \ref{h23}\right) $ (with a factor 1/2 with respect to the latter as
explained above).

The picture of the experiment in our approach is quite different from the
picture in terms of photons suggested by the HS formalism. In HS a few
photons in the (usually pulsed) laser beam are assumed to split by the
interaction with the nonlinear crystal, giving two photons each. The
probability of producing an entangled photon pair by the splitting within a
detection time is assumed of order $\left| D\right| ^{2}<<1$ whence the
simultaneous arrival of entangled photons at Alice and Bob happens for a 
\textit{small} fraction of laser pulses. However the detection of the
photons conditional to the photon production, $\eta ,$ is assumed to occur
with probability of order\textit{\ unity (}say\textit{\ }$\eta \lesssim 0.7)$%
. The probability $\eta $ is named detection efficiency.

In our aproach the probability of photocounts by Alice or Bob does not
factorize that way. Furthermore the concept of photon does not appear at
all, but there are \textit{continuous fluctuating fields including a real ZPF%
} arriving at the detectors, which are activated when the radiation
intensity is big enough.

\subsection{Understanding entanglement}

The strong correlation exhibited by the comparison of eqs.$\left( \ref{h32}%
\right) $ and $\left( \ref{h35}\right) $ is a consequence of the phenomenon
of entanglement and it is labeled strange from a classical point of view. In
our stochastic interpretation it is due to the fact that the signal field $%
Da_{i}^{*}$ produced in the crystal is correlated with the ZPF field $a_{i}$
that had enter the crystal, see eq.$\left( \ref{h27}\right) .$ Similarly for
the correlation between the signal $Da_{s}^{*}$ and the ZPF field $a_{s}$.
That is, the strong correlation appears because the same normal modes of the
radiation appear in both fields, $E_{A}$ and $E_{B}$, that go to Alice and
Bob respectively.

Now I shall stress the relevance of the vacuum fluctuations in order to
understand the difference between the ``classical correlation'' and
``entanglement''. In the evaluation of the averages eq.$\left( \ref{h}%
\right) $ we have taken the distribution of field amplitues eq.$\left( \ref
{f3}\right) $ into account giving 
\begin{equation}
\left\langle \left| a_{s}\right| ^{4}\right\rangle =\frac{1}{2}%
=2\left\langle \left| a_{s}\right| ^{2}\right\rangle ^{2},\left\langle
\left| a_{i}\right| ^{4}\right\rangle =2\left\langle \left| a_{i}\right|
^{2}\right\rangle ^{2},  \label{g1}
\end{equation}
relations typical of a Gaussian distribution of the amplitudes. Now let us
assume that we had used, instead of eq.$\left( \ref{f3}\right) $ a sure
(i.e. not fluctuating) distribution, e.g. 
\begin{equation}
\rho \left( \left\{ a_{j},a_{j}^{*}\right\} \right) =\prod_{j}\delta \left(
\left| a_{j}\right| ^{2}-\frac{1}{2}\right) ,  \label{g6}
\end{equation}
$\delta $ being Dirac delta. In this case we had obtained

\begin{equation}
\left\langle \left| a_{s}\right| ^{4}\right\rangle =\frac{1}{4}=\left\langle
\left| a_{s}\right| ^{2}\right\rangle ^{2}=\frac{1}{4}=\left\langle \left|
a_{i}\right| ^{4}\right\rangle =\left\langle \left| a_{i}\right|
^{2}\right\rangle ^{2}.  \label{g2}
\end{equation}
In this case the result for the coincidence probability had been 
\begin{equation}
P_{AB}=\frac{1}{4}\left| D\right| ^{4}=P_{A}P_{B},  \label{g3}
\end{equation}
(or $P_{AB}=0$ if we worked to order $\left| D\right| ^{2}).$ Eq.$\left( \ref
{g3}\right) $ would mean that there was no correlation between Alice and Bob
detections ! In contrast a strong positive correlation is obtained if we
take into account the fluctuations. This happens when the field is assumed
Gaussian, which leads to a stronger correlation as may be realized comparing
eq.$\left( \ref{g1}\right) $ with $\left( \ref{g2}\right) ,$ the former
leading to eqs.$\left( \ref{h35}\right) $ and the latter to $\left( \ref{g3}%
\right) $.

We conclude that the strong positive correlation associated to entanglement
requires that \textit{the fluctuations are correlated}. That is, the high
probability of coincidence detection requires a strong positive correlation
between fluctuations of the fields arriving at Alice and Bob respectively.
This leads to a physical (realistic) interpretation as follows: \textit{%
entanglement is a correlation between fluctuations\ of fields in distant
places. }In our example the correlation of fluctuations\textit{\ }involves
the vacuum fields and might be labeled entanglement between a signal and the
vacuum\cite{Milonni}, see eq.$\left( \ref{h24}\right) .$

\section{Conclusions}

\subsection{Quantum states}

A big difficulty for a realistic interpretation of quantum theory is that
the concepts of state and measurement have been highly idealized. This has
led to the attempt of achieving a picture of the quantum world, with the
(implicit) assumption that preparations and measurements are processes
simple because their mathematical representation in the HS formalism are
simple, that is vectors and self-adjoint operators. However the physical
processes involved are not simple.

Let us study the question of what is a quantum state. In classical physics
the concept of state is certainly simple, it rests on the concept of
isolation for either a particle or a wave or any combination of them.
However it is common view that neither the concept of particle nor the
concept of wave may be transferred to quantum physics. Thus the standard
answer to the question whether the electron is a particle or a wave is 
\textit{neither}. The answer involves a contradiction: anything is either
localized (particle) or extended (wave), of course with respect to some
reference size, say for an electron compared with an atom. I believe that a
more correct answer is that the electron is both. In fact an electron cannot
be seen as an isolated point particle. The physical electron corresponds to
a cloud of interacting electrons and positrons, electromagnetic radiation
and other fields with a mass\textit{\ m }and clean charge \textit{e}. The
cloud may have a size possibly as large as the Compton wavelength.

This statement may be put in a different form as follows. The vacuum
consists of a set of real fluctuating fields that are modified by the
presence of an electron. In this paper we claim that the fields are real, in
contrast with the common opinion that they are virtual. (I believe that 
\textit{virtual} is a word without any clear meaning that is used in order
to avoid commitement with either the assertion that the fields are real or
they are not). In summary, at a difference with the classical domain
particles like electrons cannot be seen as having states defined in a manner
as simple as in classical mechanics.

We may ask what is the physical interpretation of the state in a more
complex system like an atom. It is not just a system of Z+1 point (or small)
particles, that is the nucleus plus Z electrons. In the study of the atom
the nucleus might perhaps be treated as a particle localized in a region far
smaller than the atom, but this is not the case for the electrons. What
exists is a large number of electrons and positrons that are created (maybe
with emission of radiation) or annihilated (with absorption of radiation) in
pairs, with a conservation of the total electric charge, that is \textit{Ze}%
. Many other quantum fields are likely involved that may correspond to
modifications of the vacuum. In summary I believe that the quantum state of
any physical system is a quite complex structure consisting of many
interacting fields evolving in time.

Sometimes it is argued that in a nonrelativistic treatment the possible
creation or annihilation of electron-positron pairs should not be taken into
account because the energies required are far larger than typical atomic
energies. However the argument is flawed. In classical electrodynamics the
total mass-energy of say two electrons plus a positron at extremely small
distances may not be greater than the mass of a single electron due to a
possibly strong electrostatic negative energy of interaction. In summary the
internal structure of quantum systems like atoms should be always treated
taking many (relativistic) quantum fields of the vacuum into account. Of
course this is actually accepted by most people when it is recognized that
in renormalization calculations the bare mass or charge are quite different
from the physical ones. The simple change from bare to physical quantities
abridges a complicated phenomenon but the quantum formalism has the virtue
that quite complex structures like atoms may be treated using simple
equations like Schr\"{o}dinger\'{}s. That equation is just a (fairly good)
approximation. In this respect my view is quite different from the common
one. I do not believe that quantum equations are exact when we ignore the
interaction with the vacuum fields and corrections appear when the
interaction is switch on. Interactions with vacuum fields are not small
corrections, they are precisely the cause of the difference between
classical and quantum physics. Indeed classical physics is obtained from
quantum physics when $
\rlap{\protect\rule[1.1ex]{.325em}{.1ex}}h%
\rightarrow 0,$ but Planck constant $
\rlap{\protect\rule[1.1ex]{.325em}{.1ex}}h%
$ is just the parameter that fixes the scale of the vacuum fields (see
section 3), so that putting $
\rlap{\protect\rule[1.1ex]{.325em}{.1ex}}h%
=0$ means ignoring the vacuum fields.

It is remarkable that quantum theory may be formulated using simple
mathematical objects (i.e. vectors and operators in a Hilbert space) and
relations between them in order to describe complex phenomena.

\subsection{Measurements}

Measurements have been still more idealized than states in standard books or
papers on quantum mechanics. My opinions on this subject fully agree with
Einstein\'{}s. I quote him:

``You must appreciate that observation is a very complicated process. The
phenomenon under observation produces certain events in our measuring
apparatus. As a result, further processes take place in the apparatus, which
eventualIy and by complicated paths produce sense impressions and help us to
fix the effects in our consciousness. Along this whole path -from the
phenomenon to its fixation in our consciousness- we must be able to tell how
nature functions, must know the natural laws at least in practical terms,
before we can claim to have observed anything at all. Only theory, that is,
knowledge of natural laws, enables us to deduce the underlying phenomena
from our sense impressions. When we claim that we can observe something new,
we ought really to be saying that, although we are about to formulate new
natural laws that do not agree with the old ones, we nevertheless assume
that the existing laws -covering the whole path from the phenomenon to our
consciousness- function in such a way that we can rely upon them and hence
speak of observations''\cite{Heisenberg26}.

In summary a realistic interpretation of quantum theory cannot be achieved
attempting to interpret directly the (Hilbert space) formalism. That
formalism is a simple, although extremely efficient, algoritm in order to
calculate relevant predictions for the results of experiments. In some cases
alternative formalisms may be better in order to get a physical picture of
phenomena, even if they are less efficient for calculations. In particular
for the radiation field the Weyl-Wigner formalism is superior than Hilbert
space in this respect.

\end{document}